\documentclass[fleqn,usenatbib]{mnras}

\usepackage{newtxtext,newtxmath}

\usepackage[T1]{fontenc}

\DeclareRobustCommand{\VAN}[3]{#2}
\let\VANthebibliography\thebibliography
\def\thebibliography{\DeclareRobustCommand{\VAN}[3]{##3}\VANthebibliography}

\newcommand{\rb}{$R_b$}

\usepackage{graphicx}	
\usepackage{amsmath}	

\title[LMC Stars and Where to Find Them]{LMC Stars and Where to Find Them: Inferring Birth Radii for External Galaxies}

\author[Lu et al.]{Yuxi(Lucy) Lu $^{1}$\thanks{lucylulu12311@gmail.com},
          Tobias Buck$^{2,3}$,
          David Nidever$^{4}$,
          Bridget Ratcliffe$^{5}$,
          Ivan Minchev$^{5}$,
          Andrea V. Macci\`o$^{6,7,8}$,
          \newauthor
          and Aura Obreja$^{2,3}$
\\
$^{1}$American Museum of Natural History, Central Park West, Manhattan, NY, USA\\
$^{2}$Universit\"at Heidelberg, Interdisziplin\"ares Zentrum f\"ur Wissenschaftliches Rechnen, Im Neuenheimer Feld 205, Heidelberg, Germany\\
$^{3}$Universit\"at Heidelberg, Zentrum f\"ur Astronomie, Institut f\"ur Theoretische Astrophysik, Albert-Ueberle-Straße 2, Heidelberg, Germany\\
$^{4}$Montana State University, P.O. Box 173840, Bozeman, MT 59717-3840\\
$^{5}$Leibniz Institute for Astrophysics Potsdam, An der Sternwarte 16, 14482 Potsdam, Germany\\
$^{6}$New York University Abu Dhabi, PO Box 129188 Saadiyat Island, Abu Dhabi, United Arab Emirates\\
$^{7}$Center for Astrophysics and Space Science (CASS), New York University
Abu Dhabi, PO Box 129188, Abu Dhabi, UAE\\
$^{8}$Max Planck Institut f\"{u}r Astronomie, K\"{o}nigstuhl 17, D-69117 Heidelberg, Germany\\
}

\date{Accepted XXX. Received YYY; in original form ZZZ}

\pubyear{2015}

\begin{document}
\label{firstpage}
\pagerange{\pageref{firstpage}--\pageref{lastpage}}
\maketitle

\begin{abstract}
It is well known that stars move away from their birth location over time via radial migration.
This dynamical process makes computing the correct chemical evolution, e.g., metallicity gradients, of galaxies very difficult. 
This dynamical process makes inferring the chemical evolution of observed galaxies from their measured abundance gradients very difficult.
One way to account for radial migration is to infer stellar birth radii for individual stars.
Many attempts to do so have been performed over the last years, but are limited to the Milky Way as computing the birth position of stars requires precise measurements of stellar metallicity and age for individual stars that cover large Galactic radii.
Fortunately, recent and future surveys will provide numerous opportunities for inferring birth radii for external galaxies such as the Large Magellanic Cloud (LMC). 
In this paper, we investigate the possibility of doing so using the NIHAO cosmological zoom-in simulations. 
We find that it is theoretically possible to infer birth radii with a $\sim$ 25\% median uncertainty for individual stars in galaxies with i) orderliness of the orbits, $\langle v_\phi \rangle/\sigma_{v} >$ 2, ii) a dark matter halo mass greater or equal to approximately the LMC mass ($\sim$ 2 x 10$^{11} M_\odot$), and iii) after the average azimuthal velocity of the stellar disk reaches $\sim$70\% of its maximum.
From our analysis, we conclude that it is possible and useful to infer birth radii for the LMC and other external galaxies that satisfy the above criteria. 

\end{abstract}

\begin{keywords}
galaxies: abundances -- galaxies: evolution -- galaxies: kinematics and dynamics
\end{keywords}



\section{Introduction}\label{intro}
Stars can move radially in the Milky Way (MW) through heating or radial migration.
The latter process permanently changes the angular momentum of stellar orbits and occurs when stars interact with the co-rotation or overlapping resonances of the spiral arms and the bar \citep[e.g.,][]{Sellwood2002, Roskar2008, Minchev2010}.
This process is suggested by the discovery of very metal-rich stars in the solar neighborhood with small orbital eccentricities \citep{Grenon1972, Grenon1989}.
However, it was not until the early 2000s that the theory behind radial migration was developed. 
\cite{Sellwood2002} suggested that stars on horseshoe orbits around co-rotational resonances of the transient spiral arms can migrate in or out as the co-rotation locations of the spiral arms change on a timescale shorter than one full orbital period of the horseshoe orbits.
They showed that this process can create large changes in the angular momentum of the stars while keeping their orbits cold. 
Another strong mechanism for radial migration can result also from multiple long-lived asymmetric patterns, such as different spiral modes \citep{Minchev2006} or spiral and bar interaction \citep{Minchev2010}.

Since the Gaia mission \citep{gaia2016}, extensive investigations have been undertaken to gauge the impact of radial migration on the MW disk using analytic models and data \citep[e.g.,][]{Frankel2018, Frankel2019, Frankel2020, Lian2022, Wang2023, Zhang2023, Ratcliffe2023_2}. 
For example, \cite{Frankel2019} found a migration strength of 3.9 kpc $\sqrt{\tau/7\ {\rm Gyr}}$ for the past 7.5~Gyr using APOGEE \citep{APOGEE2017} red clump stars, in which $\tau$ is the look-back time.
Later, \cite{Frankel2020} suggested radial migration is one order of magnitude stronger than radial heating. 
All these studies suggest radial migration is a process that cannot be neglected in the MW. 

Simulations suggest taking into account radial migration is essential for understanding the chemical and dynamical evolution of galaxies \citep[e.g.,][]{Minchev2012, Minchev2013, Ma2017}.
Even though stars have an equal probability of migrating in and out while interacting with co-rotation resonances, the exponential stellar density profile of the Galactic disk results in more stars migrating outward than inward. 
As a result, radial abundance gradients inferred from mono-age populations without taking into account radial migration will be flatter compared to the true intrinsic gradients at the same look-back time \citep[e.g.,][]{Minchev2013, Minchev2018, Lu2022_rb, Ratcliffe2023, Anders2023}. 

The most straightforward way to take radial migration into account is to infer birth radii, \rb, for individual stars in the MW.
\cite{Minchev2018} first developed the method to infer \rb\ for individual stars in the solar neighborhood with the HARPS survey \citep{Adibekyan2012}.
They showed the potential of an empirical chemo-dynamic evolution model with this local sample of stars with precise age and metallicity measurements. 
\cite{Lu2022_rblim} later tested the assumptions in the above method and concluded that it is possible to infer \rb\ in the MW with an intrinsic uncertainty of 0.6 kpc after the stellar disk has started to form, which is believed to be $\sim$12-13 Gyr ago \citep{Conroy2022, Belokurov2022, Belokurov2023}.
Finally, \cite{Lu2022_rb} developed an empirical method to obtain the time evolution of the birth metallicity gradient using subgiant ages \citep{Xiang2022} and metallicity measurements \citep{Xiang2019} from LAMOST \citep{LAMOST}. 
Applying this method, \cite{Ratcliffe2023} directly show the birth abundance gradients and chemical evolution for stars in mono-age and mono-\rb\ populations, placing strong constraints on chemical evolution models.  

Having birth radii for external galaxies can also be advantageous as radial migration should be a universal dynamical process that can also influence the abundance gradients measured for these galaxies.
Abundance gradients measured with individual stars \citep[e.g.,][]{Povick2023, Povick2023b} or Integral Field Units \citep[IFUs; e.g.,][]{Roig2015, Ho2015, Goddard2017, Zheng2017, Parikh2021} for external galaxies can be used to test theories of galaxy formation such as different collapse models \citep[e.g.,][]{Larson1974, Spolaor2009}.
Metallicity gradient could also be an important characteristic in selecting MW-like galaxies \citep[e.g.,][]{Pilyugin2023}. 
However, the formation of metallicity gradients is complicated and can be influenced by many 
physical processes such as inside-out formation, secular dynamical effects, and external perturbations \citep[e.g.,][]{Kobayashi2004, Schroyen2013, sanchezblazquez2014, Elbadry2016}, and one way to disentangle these physical processes is to look at the relation between the metallicity gradient and mean stellar age \citep[e.g.,][]{Koleva2011, Mercado2021}. 
However, both traditionally defined radial migration caused by spiral arms \citep{Sellwood2002, Roskar2008}, bars-spiral interaction \citep{Minchev2010}, and external perturbations \citep{Quillen2009}, as well as age-dependent radial migration caused by feedback-driven outflows/inflows \citep{Elbadry2016} can bias the inference result.
For these reasons, determining birth radii for stars in external galaxies can help account for these effects, enabling us to infer the true time evolution of the abundance gradients. 

However, inferring birth radii has only been done for the MW as it requires precise measurements of metallicities and ages for individual stars that cover a large galactic radius. 
However, with current and future surveys and telescopes such as APOGEE \citep{APOGEE2017}, APOGEE-2 \citep{APOGEE2}, SDSS-V \citep{Kollmeier2017}, Roman \citep{Roman}, and JWST \citep{JWST}, we are moving into an era where observing a large number of individual stars and high spatial resolution Integral Field Units observations of external galaxies is or will be possible \citep[e.g.,][]{MaNGA, Nidever2020, gaiaLMC2021, Nidever2023, Povick2023}. 

In this paper, we use the zoom-in cosmological simulations NIHAO \citep{Wang2015} to study under which conditions it is possible to infer reasonably accurate stellar birth radii.

The paper is structured as follows: Section~\ref{sim} describes the simulations, and section~\ref{analysis} describes the statistical quantity we calculated in this work.
In section~\ref{infer}, we show that it is possible to infer \rb\ within 25\% uncertainty for galaxies with $\langle v_\phi \rangle/\sigma_{v\phi} >$ 2 and dark matter halo masses greater or equal to approximately the LMC mass \citep[$\sim$ 2 $\times$ 10$^{11} M_\odot$, ][]{Shipp2021}, and in section~\ref{import}, we show that for those galaxies for which we can infer \rb, the flattening of the true metallicity gradients at various look-back time compared to the gradients measured in mono-age populations is significant, and that \rb\ should be inferred if possible.
Finally, in section~\ref{rangefeh}, we show that it is possible to infer the time evolution of the metallicity gradient by using the range of metallicity at mono-age populations \citep[similar to what is done in][]{Lu2022_rb}. Finally, in section~\ref{lim}, we discuss some limitations of this work.

\section{Simulation \& Analysis}\label{methods}

\subsection{The NIHAO simulations}\label{sim}
NIHAO, which stands for Numerical Investigation of a Hundred Astronomical Objects, is a sample of 100 simulated galaxies with halo masses in the range $\sim 10^9-10^{12} M_\odot$. 
The dark matter halos to be re-simulated using the zoom-in method have been chosen from a large box dark matter only simulation according to an isolation criteria \citep[have no similar mass companion within less than 3 times the virial radius of the target halo at redshift $z=0$,][]{Wang2015}.

Parts of this simulation suite have been also rerun in higher resolution \citep{Buck2020} and have previously been used to study the build-up of the MW's peanut-shaped bulge \citep{Buck2018, Buck2019b}, investigate the stellar bar properties \citep{Hilmi2020}, infer the MW's dark halo spin \citep{Obreja2022}, study the dwarf galaxy inventory of MW mass galaxies \citep{Buck2019}, investigate the age-metallicity relation of MW disk stars \citep{Lu2022_rblim, Wang2023}, and the impact of early massive mergers on the metallicity gradient of the MW \citep{Buck2023}. Furthermore, these simulations have been used to study the chemical bimodality of disk stars \citep{Buck2020a,Buck2021}, their abundances \citep{Lu2022_turning}, and the origin of very metal-poor stars inside the stellar disk \citep{Sestito2021}.
Comparing the properties of these galaxies with observations of the MW and local disk galaxies from the SPARC data \citep{Lelli2016}, \citet[][]{Obreja2019} and \citet{Buck2020} showed that simulated galaxy properties agree well with observations.

\subsubsection{Simulation Prescriptions}

The NIHAO simulations assume cosmological parameters from the \cite{Planck}, namely: $\Omega_m$= 0.3175, $\Omega_\Lambda$= 0.6825, $\Omega_b$= 0.049, $H{_0}$ = 67.1 km s$^{-1}$ Mpc$^{-1}$, $\sigma_8$ = 0.8344. 
The initial conditions are created using a modified version of the \texttt{GRAFIC2} package \citep{Bertschinger2001, Penzo2014}. The resolution is such that all simulations have $\sim$10$^{\rm 6}$ particles (dark matter + gas + stellar particles) within the virial radius at $z=0$.
The mass of the particles ranges between $m_{\rm dark}\sim3.3\times10^3 - 1.7\times10^7 M_\odot$ for the dark matter 
and $m_{\rm gas}\sim6.1\times10^2 - 3.2\times10^5 M_\odot$ for the gas. 
The corresponding force softening are $\epsilon_{\rm dark}=116 - 931$ pc and $\epsilon_{\rm gas}=49 - 397$ pc.
Star particles are born with an initial mass of $1/3\times m_{\rm{gas}}$, and are subjected to mass-loss according to stellar evolution models as detailed in \citet{Stinson2013}. 
The simulation setup, star formation, and feedback implementations are described in detail in the introductory paper of the NIHAO-UHD suite \citep{Buck2020} but for completeness, we summarise them below. 

Simulations are performed with the modern smoothed particle hydrodynamics (SPH) solver {\texttt{GASOLINE2}} \citep{Wadsley2017}. 
{\texttt{GASOLINE2}} implements non-equilibrium hydrogen and helium cooling, and metal line cooling assuming photoionization equilibrium with the \citet{Haardt2005} ultraviolet background, following \citet{Shen2010}.
Star formation proceeds in cold (T $< 15,000$ K), dense ($n_{\rm  th}  >  10.3$ cm$^{-3}$) gas, and is implemented as described in \citet{Stinson2006}. 
\citet{Buck2019a} showed that with this kind of star formation model, only a high value of $n_{\rm th}>10$cm$^{-3}$ \citep[see also][for an extended parameter study]{Dutton2019, Dutton2020} can reproduce the clustering of young star clusters as observed in the Legacy Extragalactic UV Survey (LEGUS) \citep{Calzetti2015, Grasha2017}. See also \citet{Maccio2022} for a similar comparison using data from the THINGS survey \citep{Walter2008}.

As detailed in \citet{Stinson2013}, two modes of stellar feedback are implemented: (i) the energy input from young massive stars, e.g., stellar winds and photoionization, prior to the supernovae explosions, thus termed \textit{early stellar feedback} (ESF). This first mode consists of the total stellar luminosity ($2 \times 10^{50}$ erg of thermal energy per $M_{\odot}$) of the entire stellar population with an efficiency for coupling the energy input of $\epsilon_{\rm ESF}=13\%$ \citep{Wang2015}; (ii) supernova explosions implemented using the blastwave formalism as described in \citet{Stinson2006} with a delayed cooling formalism for particles inside the blast region following \citet{McKee1977} to account for the adiabatic expansion of the supernova.
Finally, we adopted a metal diffusion algorithm between particles as described in \citet{Wadsley2008}.

The halos in the zoom-in simulations were identified using the 
\texttt{Amiga Halo Finder} \citep[\texttt{AHF2},][]{Knollmann2009}, and we use the accompanying merger tree tool to trace the particle IDs of all dark matter particles through time and identify all progenitor halos of a given galaxy/dark matter halo at redshift $z=0$. Subsequent analysis of the merger tree files is then performed with the {\texttt {ytree}} package \citep{ytree}. 

\subsubsection{Scale lengths and migration strengths of NIHAO galaxies}

In this section, we calculate various physical and dynamical quantities and compare them to those of the MW and other observations.
We estimate the scale lengths, $R_d$, of the galaxies with stellar mass $M_{\rm star}> 1 \times 10^{10} M_\odot$\footnote{$M_{\rm star}$ is the total stellar mass within a 3D radius of 2.5$R_d$.} by simultaneously fitting the edge-on mass surface density profiles with a S{\'e}rsic \citep{sersic1963} and an exponential, following \cite{Buck2020}. 
For galaxies with $M_{\rm star}<$ 1$\times$10$^{10} M_\odot$, we fit only an exponential to the galaxy outskirts (75th to 90th mass weighed percentiles of stellar radii), as many did not show a clear break in the surface density profile (see also Fig.~\ref{fig:A1} in the Appendix).

We first calculate the stellar mass surface density profiles by taking the azimuthally mass-weighted average for 40-100 radial bins up to the radius enclosing 90 percent of the total stellar mass inside the virial radius.
We chose the bin size for each galaxy such that each mass surface density profile is smooth.
We then use \texttt{astropy.modeling} with a Sequential Least Squares Programming (SLSQP) optimization algorithm (\texttt{astropy.modeling.fitting.SLSQPLSQFitter}) to optimize the models and measure the scale lengths.
The stellar mass density profiles used to find the best-fitting model for each galaxy are shown in Figure~\ref{fig:A1}, and the edge-on and face-on surface density maps for the galaxies are shown in Figure~\ref{fig:1} and Figure~\ref{fig:2}.

\begin{figure}
	\includegraphics[width=\columnwidth]{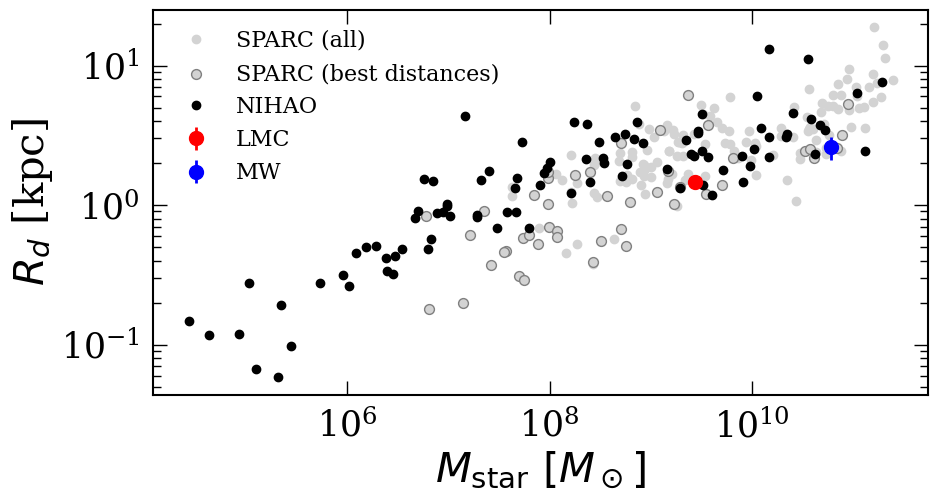}
    \caption{Galaxy scale length, $R_d$, plotted against the total stellar mass, $M_{\rm star}$, for the NIHAO simulated galaxies (black points; this work), observed external disk galaxies from Spitzer Photometry \& Accurate Rotation Curves \citep[SPARC; grey points;][]{Lelli2016}, the MW \citep[2.6 $\pm$ 0.5 kpc; blue point;][]{BlandHawthorn2016}, and the LMC \citep[1.47 $\pm$ 0.08 kpc; red point;][]{Alves2004}.
    The overall trends agree well between observations and simulations.
    The NIHAO galaxy scale length measurements are slightly higher than those from observed galaxies for galaxies with stellar mass $<$ 1$\times$10$^{9} M_\odot$.}
    \label{fig:3}
\end{figure}

Figure~\ref{fig:3} shows the scale length plotted against the total stellar mass, $M_{\rm star}$, for the NIHAO simulated galaxies (black points; this work), observed external disk galaxies from Spitzer Photometry \& Accurate Rotation Curves \citep[SPARC; grey points;][]{Lelli2016}, the MW \citep[2.6 $\pm$ 0.5 kpc; blue point;][]{BlandHawthorn2016}, and the LMC \citep[1.47 $\pm$ 0.08 kpc; red point;][]{Alves2004}.
For the observed galaxies, the stellar masses are calculated by multiplying the total luminosity at 3.6 $\mu m$ by an average mass-to-light ratio of 0.5 $M_\odot/L_\odot$. 
Compared to the observed SPARC galaxies, the scale length measurements of the NIHAO galaxies mostly agree well but appear slightly higher for galaxies with stellar mass $<$ 1$\times$10$^{9} M_\odot$.
The scale lengths of NIHAO galaxies are overall in good agreement with the observed SPARC galaxies but tend to have higher values than the observed ones at fixed stellar mass for $M_{\rm star}<$ 1$\times$10$^{9} M_\odot$.
This could be caused by the simulation prescription (e.g. resolution effects as explicitly shown in Figure A1 of \citealt{Buck2020}), or the uncertainty in the mass-to-light ratio which affects the inferred stellar mass and might shift SPARC data points left in Fig.~\ref{fig:3}.
As pointed out in \cite{Lelli2016} section 5.1, section 5.3.1, and the references within, the real average mass-to-light ratio is still unclear. 

\begin{figure}
    \includegraphics[width=\columnwidth]{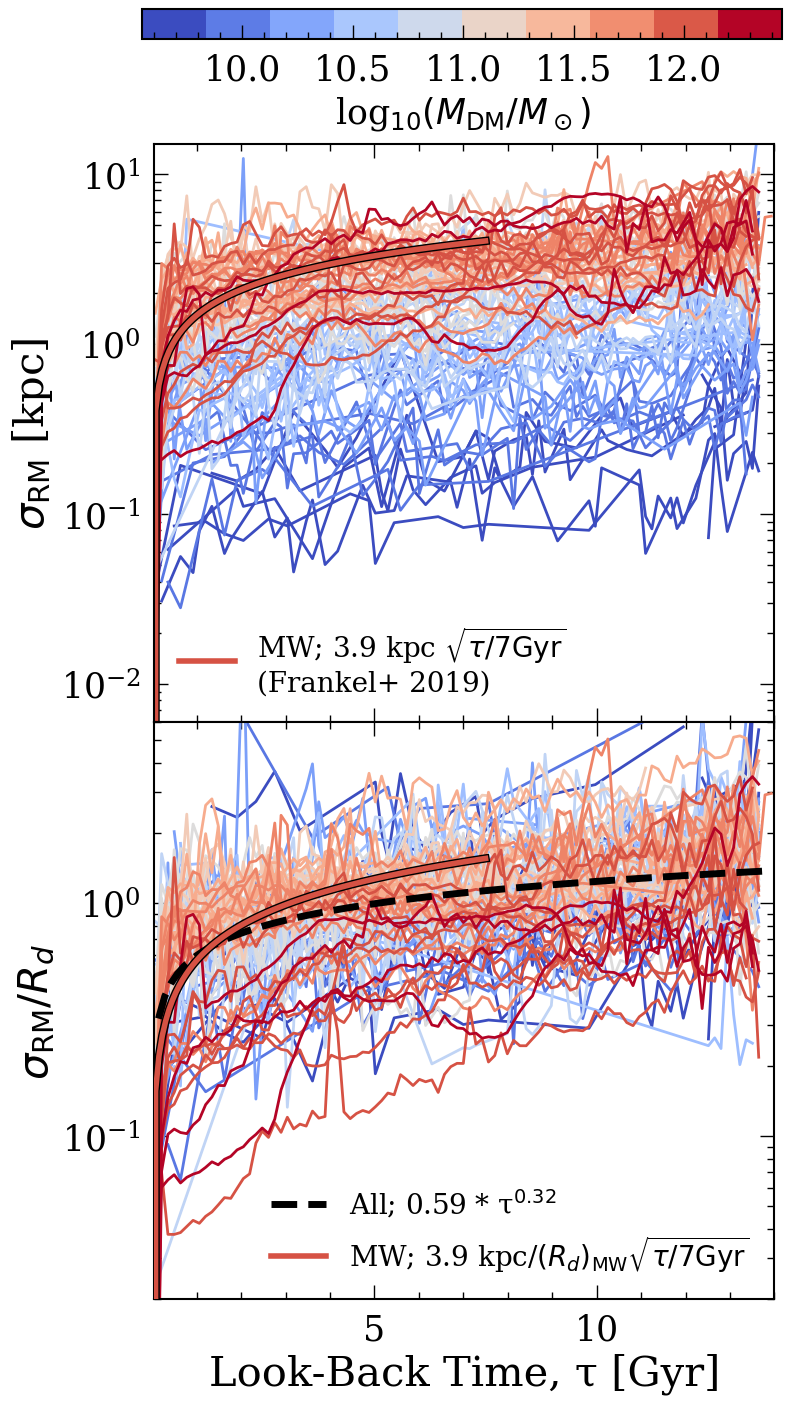}
    \caption{The radial migration strength ($\sigma_{\rm RM}$; top plot) and the normalized radial migration strength ($\sigma_{\rm RM}/R_d$; bottom plot) for each NIHAO galaxy colored by the dark matter halo mass.
    We calculated $\sigma_{\rm RM}$ by taking the standard deviation of $(R-R_b)$ in mono-age populations with a width of 0.5 Gyr.
    The bottom plot indicates radial migration is a similar gravitational process in all NIHAO galaxies.
    The red dashed lines show the analytic solution for the MW \citep[3.9 kpc $\sqrt{\tau/7 {\rm Gyr}}$, in which $\tau$ is the look-back time;][]{Frankel2019} colored by its dark matter halo mass.
    The black dashed line shows the best-fit model, indicating that radial migration is a sub-diffusion process in the NIHAO simulations.
    The scale length of the MW is taken to be 2.6 kpc \citep{BlandHawthorn2016} and its halo mass is assumed to be 10$^{\rm 12}$M$_{\rm\odot}$.
    The analytic solution fits well with the simulated galaxies of similar dark matter halo masses.}
    \label{fig:4}
\end{figure}

We also compare the radial migration strength, $\sigma_{\rm RM}$, for the simulations and the MW (see the top plot in Figure~\ref{fig:4}).
We calculate $\sigma_{\rm RM}$ by taking the standard deviation of $(R-R_b)$ in mono-age populations, same as what was done in \cite{Frankel2019}.
It is obvious that the lower-mass galaxies have significantly lower migration strengths compared to higher-mass galaxies without taking into account the scale length. 
After normalizing the migration strength by the scale length, all simulated galaxies show similar time-dependent behavior, indicating that the radial migration process is similar in most galaxies and is mostly caused by gravitational interactions. 
The thick solid lines in both plots show the analytic model for the MW taken from \cite{Frankel2019} colored by its dark matter halo mass. 
The black dashed line shows the best-fit model, indicating that radial migration is a sub-diffusion process in the NIHAO simulations.
Given $\sim$ 1$\times$10$^{12} M_\odot$ as an estimation of the dark matter halo mass of the MW, the analytic model fits well with the simulations of similar masses.

\subsection{Analysis}
\label{analysis}

According to \cite{Minchev2018}, we can infer birth radii for individual stars in a galaxy only if a strong linear gradient between [Fe/H] and \rb\ exists at look-back times when the birth radii will be inferred.
To test this assumption, we calculated the following statistical quantities for stars in mono-age bins with a width of 0.14 Gyr between 0 to 14 Gyr:
\begin{itemize}
    \item The Pearson correlation coefficient (PCC) between [Fe/H] and \rb\ for each mono-age population ---  In this work, we used the function \texttt{scipy.stats.pearsonr} to calculate the PCC. 
    \item The true birth metallicity gradient ($d{\rm [Fe/H]}/dR_b$) and the metallicity gradient inferred from mono-age populations ($d{\rm [Fe/H]}/dR$) ---
    We determined the true metallicity gradient by fitting a line to [Fe/H] vs \rb\ for stars in each mono-age population.
    This gradient represents the ISM metallicity gradient at various look-back times as stars are assumed to be born from the ISM.
    We also measured the observed metallicity gradients by fitting a line to [Fe/H] vs $R$ for stars in each mono-age population. 
    This is different from the true metallicity gradient as stars have moved away from their birth radii.
    These two gradient measurements will be used to quantify the importance of measuring birth radii in each simulation. 
\end{itemize}

\begin{figure*}
	\includegraphics[width=\textwidth]{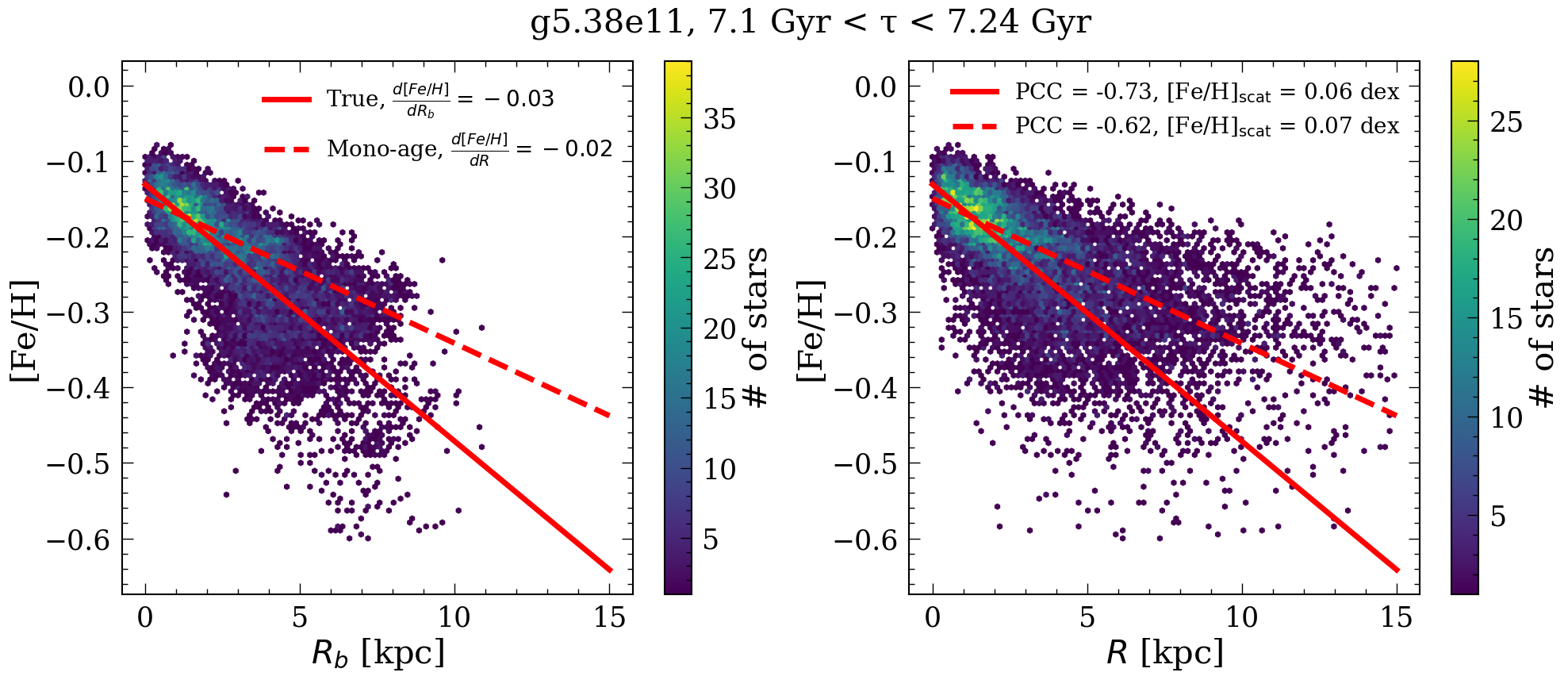}
    \caption{One example of the linear fit to obtain the Pearson correlation coefficient (PCC), the true birth metallicity gradient ($d{\rm [Fe/H]}/dR_b$), and the metallicity gradient inferred from mono-age populations ($d{\rm [Fe/H]}/dR$).
    The left plot shows the [Fe/H]-\rb\ relation, and the right plot shows the [Fe/H]-$R$ relation.
    Not only is the [Fe/H]-$R$ calculated from a small mono-age bin significantly flatter than the true metallicity gradient at birth, but the scatter is also 15\% higher.
    This figure further suggests that not taking radial migration into account will result in an incorrect result for the time evolution of the true metallicity gradient.}
    \label{fig:5}
\end{figure*}

Figure~\ref{fig:5} shows one example of the fit to calculate PCC, $d{\rm [Fe/H]}/dR_b$, and $d{\rm [Fe/H]}/dR$ for stars between 7.1-7.24 Gyr in the \texttt{g5.38e11} galaxy. 
In this case, the gradient in the mono-age population is significantly (33\%) flatter than that of the true birth gradient. 

Besides the properties we measured for stars of mono-age populations, we also measured the following global properties and summary statistics for the NIHAO galaxies.
\begin{itemize}
    \item The intrinsic median fractional uncertainty in measuring \rb\ for each galaxy. 
    We calculated this by first inferring \rb\ for individual stars using the method described in \cite{Minchev2018}.
    This estimates the \rb\ uncertainty assuming the birth metallicity gradient is linear at all times and there is no scatter around the line. 
    In detail, we first measure the time evolution of the ISM metallicity by fitting 1-D polynomials (lines) to [Fe/H] and \rb\ for stars in mono-age bins (e.g., the solid line in Figure~\ref{fig:5}).
    We then infer \rb\ for each star from its metallicity and age based on the time evolution of the ISM metallicity measured from the previous step.
    Finally, we calculate the median intrinsic uncertainty using the median absolute deviation, given by: median($|(R_b)_{\rm inf}-(R_b)_{\rm true}|/(R_b)_{\rm true}$), in which $(R_b)_{\rm inf}$ is the inferred \rb, assuming [Fe/H]-\rb\ has a perfectly linear relation, and $(R_b)_{\rm true}$ is the true \rb. 
    This median uncertainty quantifies how well we can infer birth radii if the time evolution of the ISM metallicity is known, as also done in \cite{Lu2022_rblim}.
    \item The orderliness of the stellar orbits ($\langle v_\phi \rangle/\sigma_{v}$).
    We calculate the orderliness of the stellar orbit by taking the average azimuthal velocity $v_\phi$ at 2.2$R_d$ and dividing it by the total velocity dispersion ($\sigma_v$) for stars at the same radius. 
    This quantity describes the ``diskyness'' of the galaxy, e.g., disk galaxies typically have $\langle v_\phi \rangle/\sigma_{v} >$ 3.
\end{itemize}

\section{Results}\label{result}

\subsection{Criteria to infer birth radii for external galaxies}\label{infer}
\begin{figure*}
	\includegraphics[width=\textwidth]{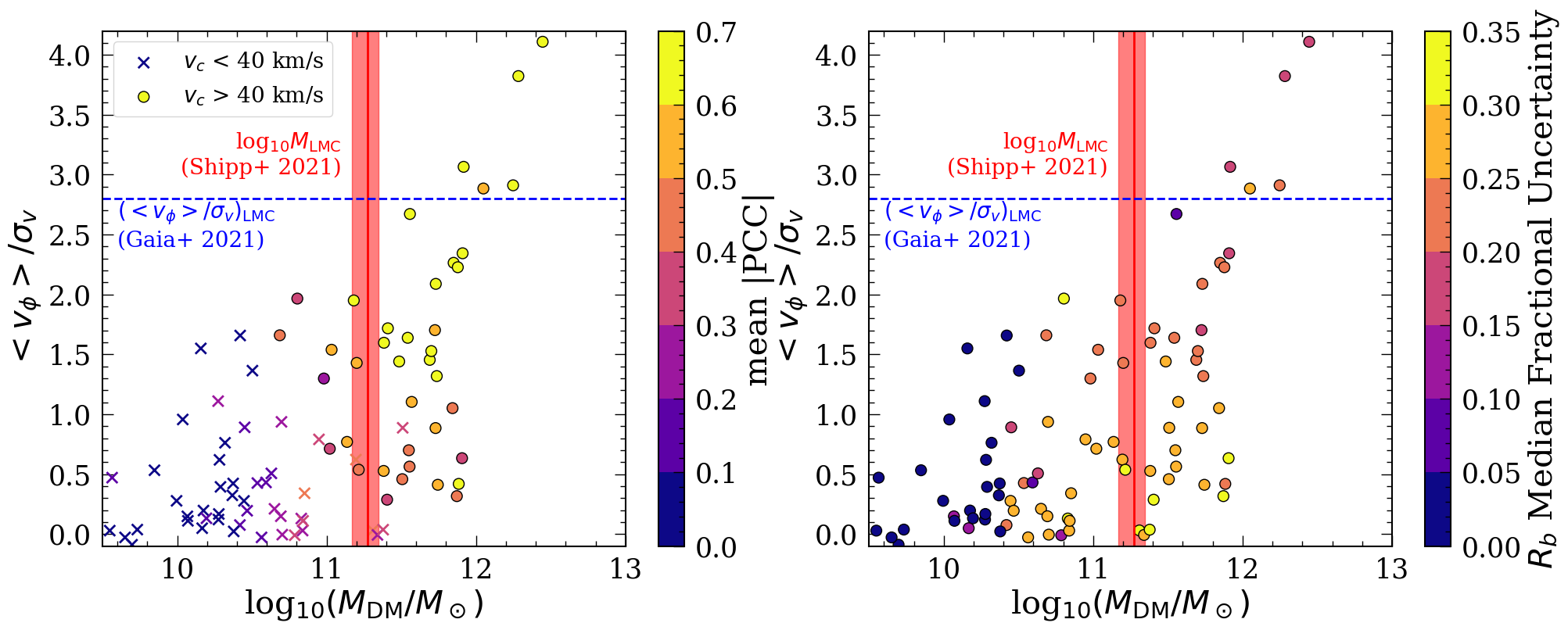}
    \caption{$\langle v_\phi \rangle/\sigma_{v}$ versus the dark matter halo mass colored by the mean absolute PCC (left) and the median fractional intrinsic uncertainty of measuring \rb\ (right).
    The crosses and circles in the left plot indicate whether the flat rotation curve measured at 2.2$R_d$ is greater than 40 km/s or not \citep[galaxies with flat rotation curve $<$ 40 km/s form as pressure-supported systems that are thick and puffy][]{Kaufmann2007}.
    The vertical red line in the two panels gives the estimated LMC halo mass \citep[18.8$^{+3.5}_{-4.0}$$\times$10$^{10} M_\odot$;][]{Shipp2021}.
    The horizontal blue dashed line in both panels shows the estimated LMC's diskyness $\langle v_\phi \rangle/\sigma_{v}\sim$~2.8 \citep[Figure B.5. in][]{gaiaLMC2021}.
    Based on the measurements for the LMC, we concluded that we are theoretically able to infer \rb\ within $\sim$ 25\% uncertainty.}
    \label{fig:6}
\end{figure*}

The definition of birth radius requires the galaxy to have a preferred rotation direction.
As a result, both the mass and the morphology of the galaxy should be important in dictating whether or not birth radii can be inferred. 
The left panel of Figure~\ref{fig:6} shows $\langle v_\phi \rangle/\sigma_{v}$ versus the dark matter halo mass, colored by the mean absolute PCC, which is the absolute value of the PCC averaged over all look-back times.
The right panel of Figure~\ref{fig:6} shows the same data but colored by the median fractional intrinsic uncertainty of measuring \rb.
The red lines and the shaded areas show the estimation for the LMC mass \citep[18.8$^{+3.5}_{-4.0}$$\times$10$^{10} M_\odot$;][]{Shipp2021}.
The blue dashed lines show $\langle v_\phi \rangle/\sigma_{v}$ for the LMC, estimated to be $\sim$ 2.8 using 
combined stellar population \citep[Figure B.5. in][]{gaiaLMC2021}. 

Going from low-mass dwarfs to higher-mass galaxies, the average |PCC| increases.
This is expected as lower mass systems are mainly pressure-supported, forming stars less efficiently and on heated orbits.
On the contrary, larger galaxies are more likely to form thin, cool disks that are rotationally supported.
According to \cite{Kaufmann2007}, galaxies with a flat rotation curve $< 40$ km/s do not form thin disks.
This agrees with the NIHAO galaxies, in that galaxies with flat rotation curves $<$ 40 km/s at 2.2$R_d$ have $\langle v_\phi \rangle/\sigma_{v} \lesssim1.5$ and average |PCC| $< 0.6$ (Figure~\ref{fig:6} left plot).  
This means a flat rotation curve greater than 40 km/s is a basic requirement to be able to infer birth radii. 
In the NIHAO sample, a strong linear metallicity gradient with |PCC| $>$ 0.7 only exists in galaxies with $\langle v_\phi \rangle/\sigma_{v} \gtrsim1.5$ and dark matter halo mass greater than the LMC's mass.
However, a strong linear gradient does not guarantee a low uncertainty in inferring \rb.
To have a low uncertainty in \rb, we also need a steep metallicity gradient. 
Fortunately, more rotationally supported galaxies also have steeper gradients on average \citep[e.g.,][]{Goddard2017, Nanni2023}, and thus, have smaller median intrinsic uncertainty in inferring \rb\ (See Figure~\ref{fig:6} right plot). 

\begin{figure}
    \includegraphics[width=\columnwidth]{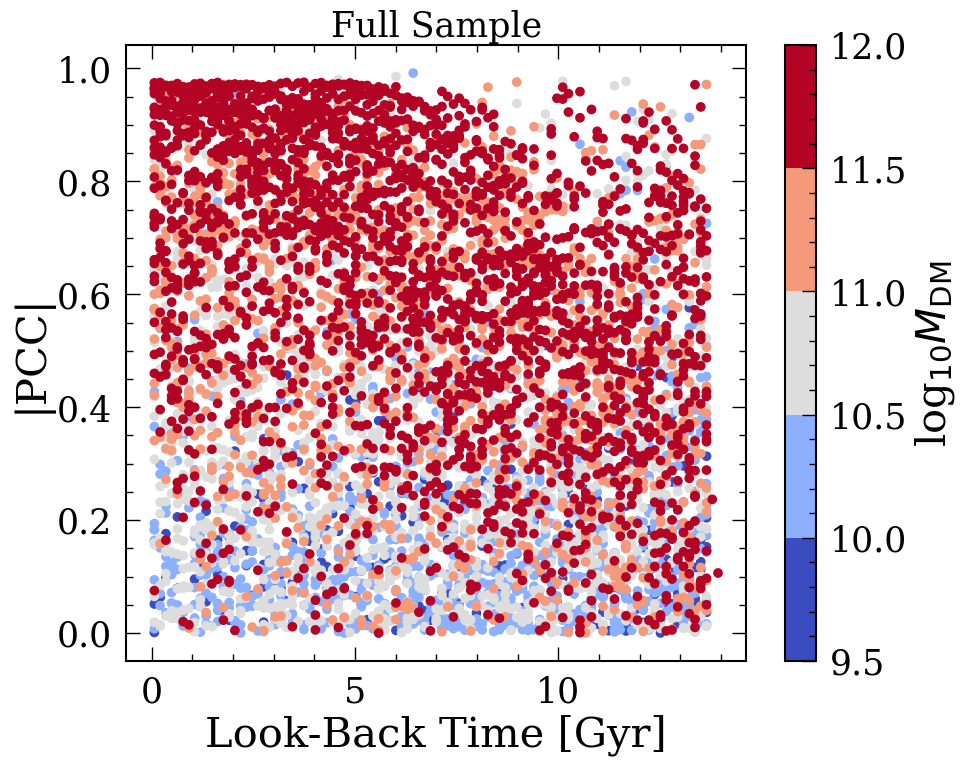}
	\includegraphics[width=0.9\columnwidth]{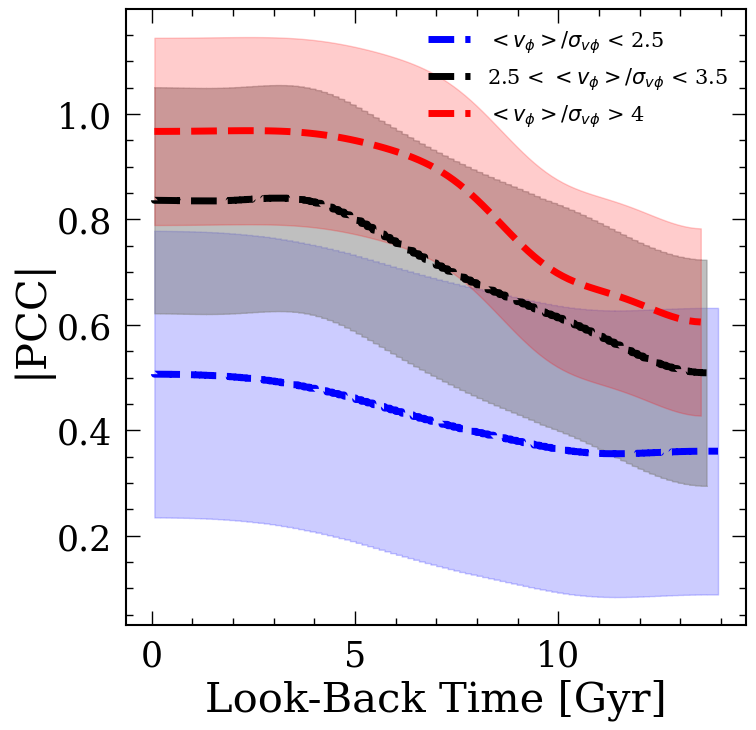}
    \includegraphics[width=0.91\columnwidth]{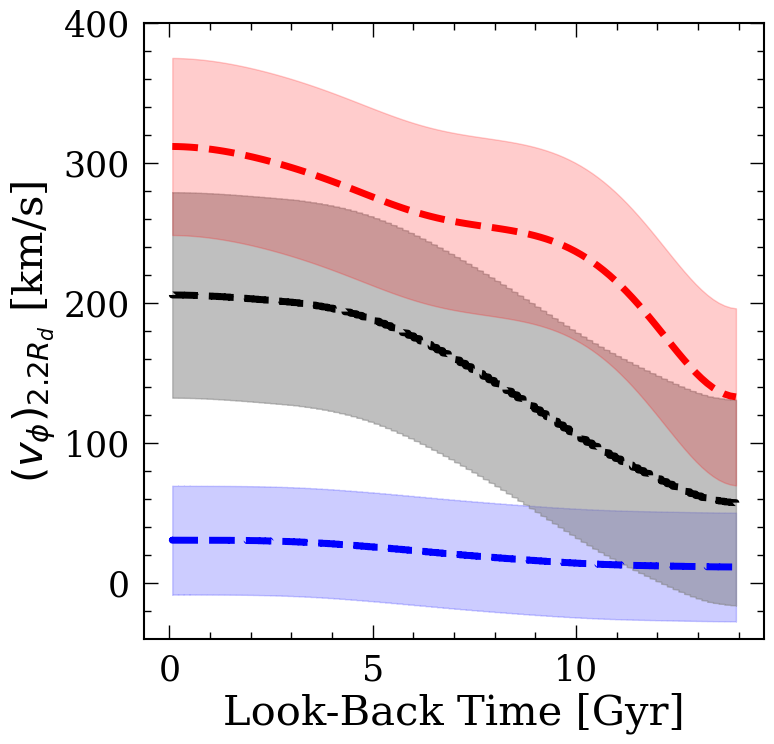}
    \caption{Top: |PCC| of [Fe/H]-\rb\ versus look-back time for each NIHAO galaxy colored by their dark matter halo mass. 
    Middle: mean and standard deviation of |PCC| versus look-back time for three different $\langle v_\phi \rangle/\sigma_{v}$ bins.
    Bottom: same as middle but for $(v_\phi)_{2.2R_d}$.}
    \label{fig:7}
\end{figure}

In all the previous analyses, we marginalized over the time axis.
Now, we look at the temporal evolution of the |PCC| and $(v_\phi)_{2.2R_d}$ as a function of dark matter halo mass.
The top panel of Figure~\ref{fig:7} shows the |PCC| for all galaxies as a function of look-back time, colored by their dark matter halo mass. 
It is clear that the |PCC| measurements for galaxies with higher dark matter halo mass are on average, higher than those for lower masses at all times, and that a linear relation (|PCC| $>$ 0.7) between [Fe/H] and \rb\ is not well established before a look-back time of $\sim$ 10 Gyr, around the time when the stellar disks have started to form for the higher mass galaxies \citep[see also][]{Lu2022_rblim}. 
The red, black, and blue dashed lines and shaded areas in the middle and bottom panels of Figure~\ref{fig:7} show the mean and standard deviation of the |PCC| (middle) and $(v_\phi)_{2.2R_d}$ (bottom) for three different bins of $\langle v_\phi \rangle/\sigma_{v}$ indicated in the legends.
It is even more clear from these two plots that larger galaxies not only form stronger linear gradients, they also form them earlier, around the time that the stellar disks have started to form, which is in agreement with the derived \rb\ evolution for the MW in the works by \cite{Minchev2018} and \cite{Lu2022_rb}.
It is also worth pointing out that the ``spin-up'' of the disk, which indicates the time when the stellar disk has started to form, occurs at a look-back time of $\sim$12-13 Gyr for the simulated galaxies that are the most disky ($\langle v_\phi \rangle/\sigma_{v} >$ 4; red dashed line), agreeing remarkably well with the MW observations \citep{Conroy2022, Belokurov2022, Belokurov2023}.

\begin{figure}
	\includegraphics[width=\columnwidth]{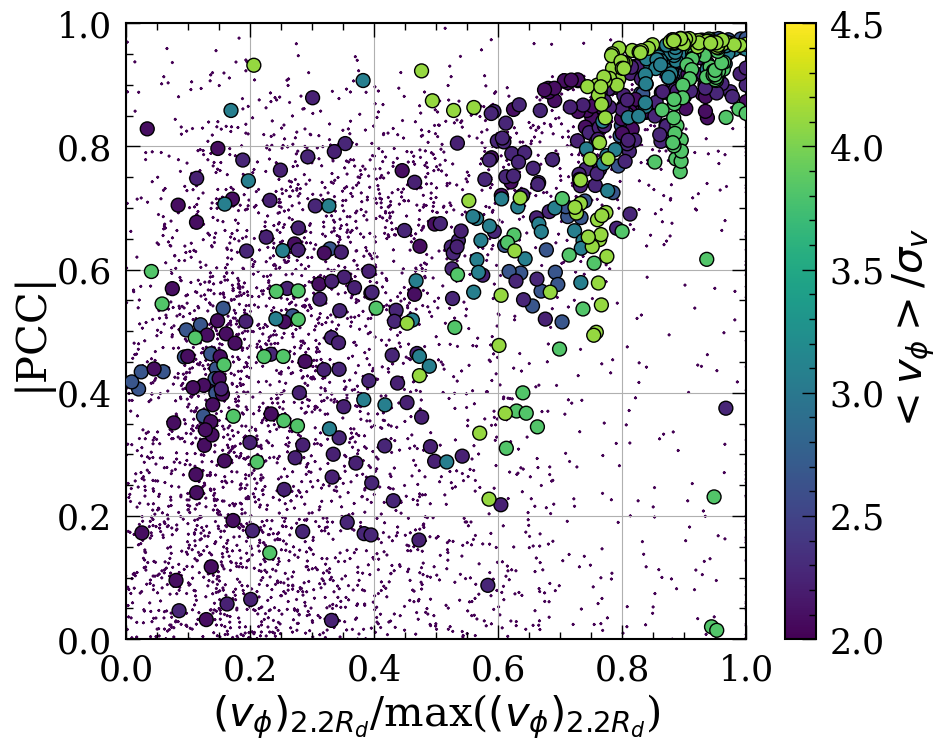}
    \caption{|PCC| of [Fe/H]-\rb\ versus $(v_\phi)_{2.2R_d}$/max($(v_\phi)_{2.2R_d}$ for stars in every mono-age populations with $\langle v_\phi \rangle/\sigma_{v} < 2$ (dots) and $\langle v_\phi \rangle/\sigma_{v} > 2$ (points).
    The points and dots are colored by the orderliness of the stellar orbits. 
    Inferring \rb\ with an average intrinsic uncertainty of 25\% can be done once the average azimuthal velocity of the stellar disk reaches $\sim$70\% of the maximum.}
    \label{fig:8}
\end{figure}

Figure~\ref{fig:8} plots the |PCC| versus $(v_\phi)_{2.2R_d}$/max($(v_\phi)_{2.2R_d}$) for stars in mono-age populations.
$(v_\phi)_{2.2R_d}$/max($(v_\phi)_{2.2R_d}$ -- calculated by normalizing $(v_\phi)_{2.2R_d}$ at each look-back time with the maximum of all measured $(v_\phi)_{2.2R_d}$ -- indicates how established the stellar disk is.
A value close or equal to 1 means the stellar disk is fully established and normally occurs at a later look-back time as $(v_\phi)_{2.2R_d}$ gradually increases with time (see Figuer~\ref{fig:7}).
Interestingly enough, for more rotationally supported galaxies ($\langle v_\phi \rangle/\sigma_{v} > 2$), a strong linear correlation exist between [Fe/H] and \rb\ (|PCC| $>$ 0.7) when $(v_\phi)_{2.2R_d}$/max($(v_\phi)_{2.2R_d}$) reaches $\sim$ 0.7. 
Since $(v_\phi)_{2.2R_d}$/max($(v_\phi)_{2.2R_d}$) increases as stars get younger, this means for the NIHAO galaxies, \rb\ can be inferred with an average intrinsic uncertainty $\sim$ 25\% for stars formed after a look-back time when the average azimuthal velocity in the rotating disk is 70\% of the youngest stars.
This could potentially be used to identify the age range of stars for which we can infer reliable \rb\ for external galaxies. 
However, it is worth pointing out that the scatter is large until the average azimuthal velocity of the stellar disk is $\sim$80\% of that of the youngest stars. 
As a result, without analyzing individual galaxies, one should only infer \rb\ for stars younger than the age when the average azimuthal velocity of the stellar disk is $\sim$80\% of that of the youngest stars.

\subsection{Importance of radial migration in inferring the birth metallicity gradient}
\label{import}

One important question, before actually inferring \rb\ for any external galaxies, is to understand whether it is a useful quantity. 
One way to answer this question is to look at the differences between the true metallicity gradients at various look-back times ($d$[Fe/H]/$d$\rb) and the metallicity gradients measured from mono-age populations ($d$[Fe/H]/$dR$). 
If the difference between the two is significant, being able to infer \rb\ will mean that we can understand the true time evolution of the metallicity gradient, and hence constrain how the galaxy formed.

\begin{figure}
	\includegraphics[width=\columnwidth]{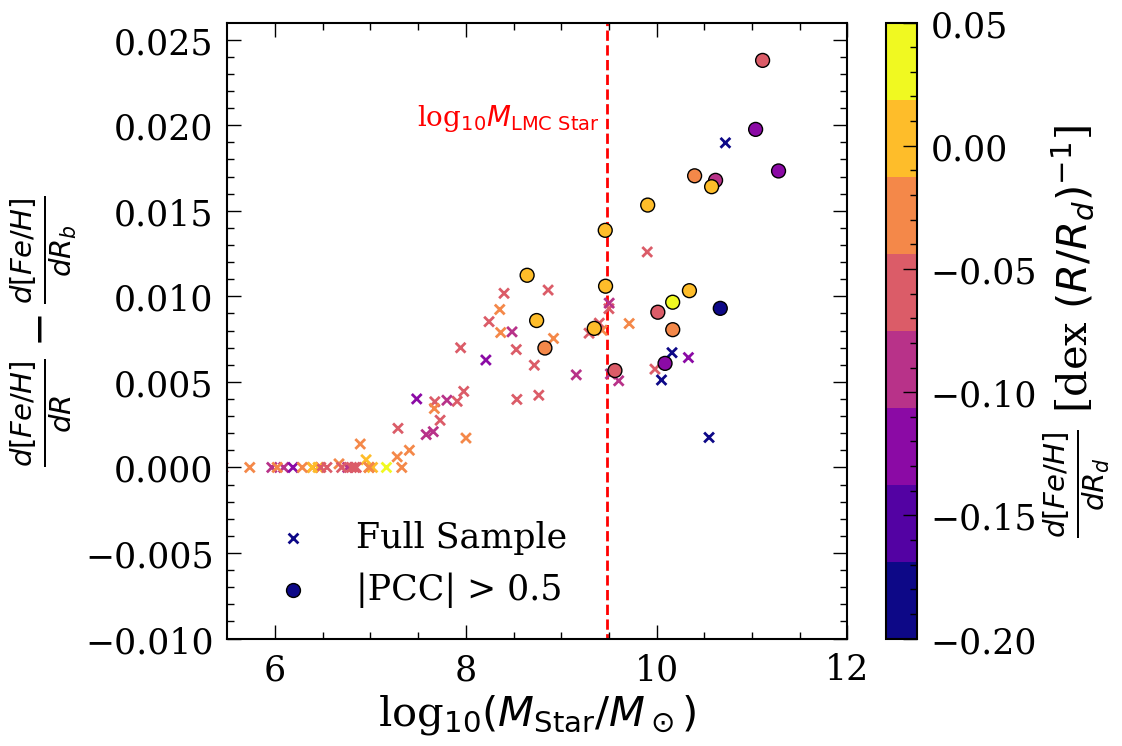}
    \caption{Average difference between the gradient measured from mono-age populations and the true birth gradient. 
    The color shows the overall gradient measurement using all the stars today, normalized by the galaxy scale length.
    The points show the galaxies with an average |PCC| $>$ 0.6 and an average intrinsic uncertainties in \rb\ $<$ 25\%.
    The crosses show the rest of the galaxies. 
    The red vertical dashed line shows the estimation of the LMC stellar mass \citep[$\sim$3$\times$10$^{9} M_\odot$;][]{vanderMarel2014}.
    For galaxies for which we can infer reliable \rb, the gradients measured from mono-age populations are on average 0.005 dex/kpc or $\sim$ 30\% flatter than the true metallicity gradients.
    This figure suggests that the ability to infer birth radii holds significant value. }
    \label{fig:9}
\end{figure}

Figure~\ref{fig:9} shows the average differences between the metallicity gradient measured from mono-age populations and the true birth gradient. 
The color represents the metallicity gradient measurement at redshift 0, normalized by the galaxy scale length.
The points show the galaxies with an average |PCC| $>$ 0.6 and a median intrinsic uncertainties in \rb\ $<$ 25\%.
The crosses show the rest of the galaxies. 
The red vertical dashed line shows the estimation of the LMC stellar mass \citep[$\sim$3$\times$10$^{9} M_\odot$;][]{vanderMarel2014}.

As expected, the metallicity gradient flattens over time, indicated by the positive values of the mean differences between the gradients on the $y$-axis.
Moreover, more massive galaxies experience more flattening in general, as they have higher migration strengths compared to lower mass galaxies (see Figure~\ref{fig:4} top plot). 
There is a clear transition around 
stellar masses of $\sim$ 10$^{8} M_\odot$, which corresponds to a dark matter halo mass $\sim$ 3$\times$10$^{10} M_\odot$.
This is around the mass where the galaxies transition between more pressure supported to rationally supported ($(v_\phi)_{2.2R_d} >$ 40 km/s; see Figure~\ref{fig:6}).
No strong correlation between the present day metallicity gradient normalized by the galaxy size (color of the points) and stellar mass is seen for galaxies with stellar mass $<$ 10$^{10} M_\odot$. 
For those galaxies for which we can infer birth radii, the metallicity gradients measured from mono-age populations are significantly flatter than the true metallicity gradients at the same look-back time, suggesting the ability to infer birth radii hold important value for any galaxy for which we can do so. 

\section{Discussion}
\label{disc}

\subsection{Relating current observables with the metallicity gradient at birth}
\label{rangefeh}

\begin{figure}
	\includegraphics[width=\columnwidth]{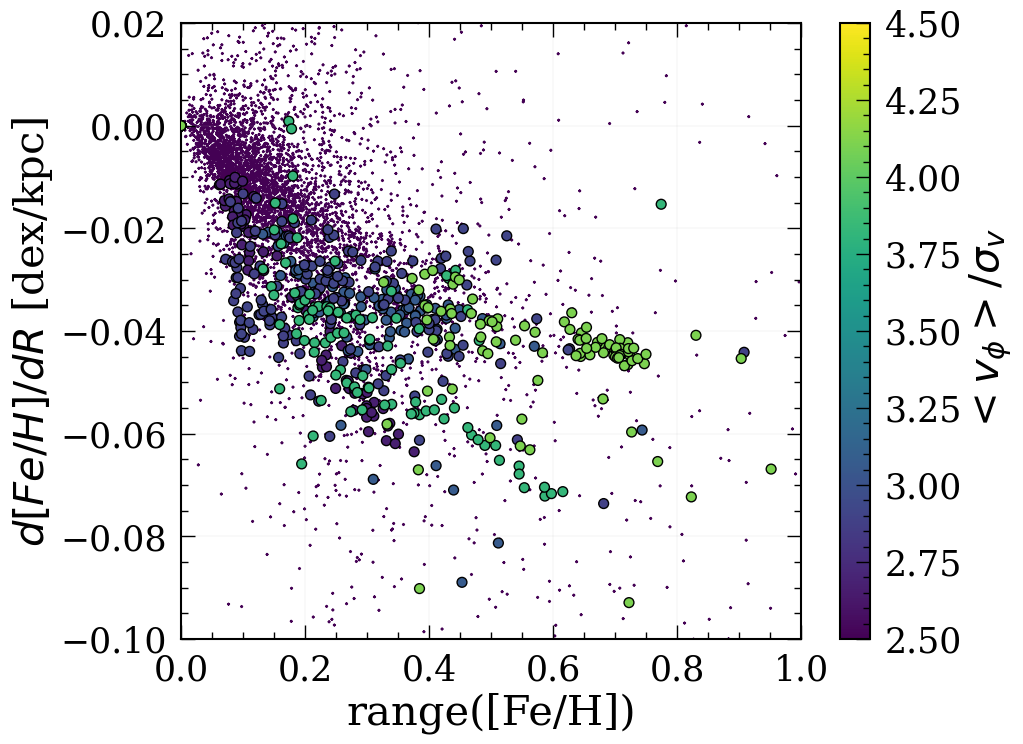}
    \caption{Relation between the metallicity gradient at each look-back time and the range in metallicity (defined to be [Fe/H]$_{\rm 95\%}$ - [Fe/H]$_{\rm 5\%}$) for the corresponding mono-age population \citep[similar to Figure A1 in][for each NIHAO galaxy]{Lu2022_rb}. 
    The points and dots show galaxies with $\langle v_\phi \rangle/\sigma_{v} > 2$ and $\langle v_\phi \rangle/\sigma_{v} < 2$, respectively.
    It is clear that strong anti-correlations exist between the measurements of the range in metallicity in mono-age populations and the metallicity gradients at corresponding look-back times.}
    \label{fig:10}
\end{figure}

\cite{Lu2022_rb} recently developed a way to empirically obtain the time evolution of the metallicity gradient by noting that the metallicity gradient at each look-back time scales as the inverse of the range in metallicity [Fe/H] of the corresponding mono-age population.
Figure~\ref{fig:10} shows the relation between the metallicity gradient at each look-back time and the metallicity range for the corresponding mono-age population \citep[similar to Figure A1 in][for each NIHAO galaxy]{Lu2022_rb}. 
The metallicity range is calculated by taking the 95\% percentile - 5\% percentile in metallicity for the star particles, the same as what was done in \cite{Lu2022_rb}.
The points and dots show galaxies with $\langle v_\phi \rangle/\sigma_{v} > 2$ and $\langle v_\phi \rangle/\sigma_{v} < 2$, respectively. 
It is clear that there exists a strong anti-correlation between the range in metallicity in mono-age populations and the metallicity gradient at the corresponding look-back time.
This figure serves as a proof-of-concept for future work to obtain empirical birth radii for the LMC and beyond using the method developed in \cite{Lu2022_rb}, which has been shown to work for the TNG50 simulated galaxies \citep{Pillepich2019, Nelson2019a_TNG, Nelson2019} from the IllustrisTNG project \citep{Pillepich2018_TNG, Marinacci2018_TNG, Naiman2018_TNG, Springel2018_TNG, Nelson2018_TNG, Nelson2019a_TNG} with higher, MW and M31-like masses (Ratcliffe et al. in prep).

\subsection{Limitations}\label{lim}
It is worth noting that this work is only based on one suite of simulations, and while simulations in the cosmological context provide important insights into galaxy formation, unknown subgrid physics, limitations in resolution, different feedback mechanisms, and chemical enrichment prescriptions (e.g., yields) may affect our results.
Moreover, we did not take into account the effects of measurement uncertainties in stellar age and metallicity. 

Finally, we did not study the effects of mergers, star formation rates, and galaxy morphology.
The scatter seen in each figure is likely caused by the differences in the detailed formation scenario for each galaxy. 
Future work should conduct a more detailed study of individual galaxies.

\section{Conclusions}
Key summary results of the paper are:
\begin{itemize}
    \item One can theoretically infer reliable \rb\ within an average intrinsic uncertainty of $\sim$ 25\% for 
    \begin{enumerate}
        \item Galaxies with dark matter halo mass $>$ LMC mass \citep[$\sim$ 2$\times$10$^{11}M_\odot$;][]{Shipp2021}, and the an orderliness of the stellar orbits as quantified by, $\langle v_\phi \rangle/\sigma_{v}$ $>$ 2 (see Figure~\ref{fig:6}).
        \item Stars that are younger than those of a look-back time when the azimuthal velocity of the stellar disk reaches $\sim$70\% of the maximum (see Figure~\ref{fig:7} and Figure~\ref{fig:8}).
    \end{enumerate}
    \item For those galaxies for which we can infer birth radii, the metallicity gradients measured from mono-age populations are significantly flatter than the true metallicity gradients at the same look-back times (see Figure~\ref{fig:9}).
    \item And finally, our main conclusion: It is possible to infer birth radii for individual stars in the LMC with $\sim$ 25\% uncertainty (Figure~\ref{fig:6}) given precise and accurate age and metallicity measurements.
    This means, we are theoretically able to put stars back to their birth radii and recover the true chemical evolution of the LMC. 
\end{itemize}

\section*{Acknowledgements}
Y.L. thanks for the helpful discussions provided by the MW \& stars meeting at Columbia University and the Local Volume group meeting at the Center for Computational Astrophysics. T.B.'s and A.O.'s contributions to this project were made possible by funding from the Carl Zeiss Foundation. We gratefully acknowledge the Gauss Centre for Supercomputing e.V. (www.gauss-centre.eu) for funding this project by providing computing time on the GCS Supercomputer SuperMUC at Leibniz Supercomputing Centre (www.lrz.de). I.M. and B.R. acknowl- edge support by the Deutsche Forschungsgemeinschaft under the grant MI 2009/2-1. This research was carried out on the High Performance Computing resources at New York University Abu Dhabi.
This research made use of the {\sc{pynbody}} \citet{pynbody} package to analyze the simulations and used the {\sc{ytree}} package \citep{ytree} to analyze the AHF merger trees. We made use of the {\sc{python}} package {\sc{matplotlib}} \citep{matplotlib} to display all figures in this work. Data analysis for this work made intensive use of the {\sc{python}} library {\sc{SciPy}} \citep{scipy}, in particular {\sc{NumPy, IPython and Jupyter notebooks}} \citep{numpy,ipython,jupyter}.

\section*{Data Availability}
The NIHAO simulations underlying this work are available upon reasonable request to the authors. Redshift zero snapshots of a few high-resolution versions from the NIHAO-UHD project are publicly available at \url{https://tobias-buck.de/#sim_data}.



\bibliographystyle{mnras}
\bibliography{example} 

\begin{thebibliography}{}
\makeatletter
\relax
\def\mn@urlcharsother{\let\do\@makeother \do\$\do\&\do\#\do\^\do\_\do\%\do\~}
\def\mn@doi{\begingroup\mn@urlcharsother \@ifnextchar [ {\mn@doi@} {\mn@doi@[]}}
\def\mn@doi@[#1]#2{\def\@tempa{#1}\ifx\@tempa\@empty \href {http://dx.doi.org/#2} {doi:#2}\else \href {http://dx.doi.org/#2} {#1}\fi \endgroup}
\def\mn@eprint#1#2{\mn@eprint@#1:#2::\@nil}
\def\mn@eprint@arXiv#1{\href {http://arxiv.org/abs/#1} {{\tt arXiv:#1}}}
\def\mn@eprint@dblp#1{\href {http://dblp.uni-trier.de/rec/bibtex/#1.xml} {dblp:#1}}
\def\mn@eprint@#1:#2:#3:#4\@nil{\def\@tempa {#1}\def\@tempb {#2}\def\@tempc {#3}\ifx \@tempc \@empty \let \@tempc \@tempb \let \@tempb \@tempa \fi \ifx \@tempb \@empty \def\@tempb {arXiv}\fi \@ifundefined {mn@eprint@\@tempb}{\@tempb:\@tempc}{\expandafter \expandafter \csname mn@eprint@\@tempb\endcsname \expandafter{\@tempc}}}

\bibitem[\protect\citeauthoryear{{Abdurro'uf} et~al.,}{{Abdurro'uf} et~al.}{2022}]{APOGEE2}
{Abdurro'uf} et~al., 2022, \mn@doi [\apjs] {10.3847/1538-4365/ac4414}, \href {https://ui.adsabs.harvard.edu/abs/2022ApJS..259...35A} {259, 35}

\bibitem[\protect\citeauthoryear{{Adibekyan}, {Sousa}, {Santos}, {Delgado Mena}, {Gonz{\'a}lez Hern{\'a}ndez}, {Israelian}, {Mayor}  \& {Khachatryan}}{{Adibekyan} et~al.}{2012}]{Adibekyan2012}
{Adibekyan} V.~Z.,  {Sousa} S.~G.,  {Santos} N.~C.,  {Delgado Mena} E.,  {Gonz{\'a}lez Hern{\'a}ndez} J.~I.,  {Israelian} G.,  {Mayor} M.,   {Khachatryan} G.,  2012, \mn@doi [\aap] {10.1051/0004-6361/201219401}, \href {https://ui.adsabs.harvard.edu/abs/2012A&A...545A..32A} {545, A32}

\bibitem[\protect\citeauthoryear{{Alves}}{{Alves}}{2004}]{Alves2004}
{Alves} D.~R.,  2004, \mn@doi [\apjl] {10.1086/382130}, \href {https://ui.adsabs.harvard.edu/abs/2004ApJ...601L.151A} {601, L151}

\bibitem[\protect\citeauthoryear{{Anders} et~al.,}{{Anders} et~al.}{2023}]{Anders2023}
{Anders} F.,  et~al., 2023, \mn@doi [arXiv e-prints] {10.48550/arXiv.2304.08276}, \href {https://ui.adsabs.harvard.edu/abs/2023arXiv230408276A} {p. arXiv:2304.08276}

\bibitem[\protect\citeauthoryear{{Belokurov} \& {Kravtsov}}{{Belokurov} \& {Kravtsov}}{2022}]{Belokurov2022}
{Belokurov} V.,  {Kravtsov} A.,  2022, \mn@doi [\mnras] {10.1093/mnras/stac1267}, \href {https://ui.adsabs.harvard.edu/abs/2022MNRAS.514..689B} {514, 689}

\bibitem[\protect\citeauthoryear{{Belokurov} \& {Kravtsov}}{{Belokurov} \& {Kravtsov}}{2023}]{Belokurov2023}
{Belokurov} V.,  {Kravtsov} A.,  2023, \mn@doi [arXiv e-prints] {10.48550/arXiv.2309.15902}, \href {https://ui.adsabs.harvard.edu/abs/2023arXiv230915902B} {p. arXiv:2309.15902}

\bibitem[\protect\citeauthoryear{{Bertschinger}}{{Bertschinger}}{2001}]{Bertschinger2001}
{Bertschinger} E.,  2001, \mn@doi [\apjs] {10.1086/322526}, \href {http://adsabs.harvard.edu/abs/2001ApJS..137....1B} {137, 1}

\bibitem[\protect\citeauthoryear{{Bland-Hawthorn} \& {Gerhard}}{{Bland-Hawthorn} \& {Gerhard}}{2016}]{BlandHawthorn2016}
{Bland-Hawthorn} J.,  {Gerhard} O.,  2016, \mn@doi [\araa] {10.1146/annurev-astro-081915-023441}, \href {https://ui.adsabs.harvard.edu/abs/2016ARA&A..54..529B} {54, 529}

\bibitem[\protect\citeauthoryear{{Buck}}{{Buck}}{2020}]{Buck2020a}
{Buck} T.,  2020, \mn@doi [\mnras] {10.1093/mnras/stz3289}, \href {https://ui.adsabs.harvard.edu/abs/2020MNRAS.491.5435B} {491, 5435}

\bibitem[\protect\citeauthoryear{{Buck}, {Ness}, {Macci{\`o}}, {Obreja}  \& {Dutton}}{{Buck} et~al.}{2018}]{Buck2018}
{Buck} T.,  {Ness} M.~K.,  {Macci{\`o}} A.~V.,  {Obreja} A.,   {Dutton} A.~A.,  2018, \mn@doi [\apj] {10.3847/1538-4357/aac890}, \href {https://ui.adsabs.harvard.edu/abs/2018ApJ...861...88B} {861, 88}

\bibitem[\protect\citeauthoryear{{Buck}, {Macci{\`o}}, {Dutton}, {Obreja}  \& {Frings}}{{Buck} et~al.}{2019a}]{Buck2019}
{Buck} T.,  {Macci{\`o}} A.~V.,  {Dutton} A.~A.,  {Obreja} A.,   {Frings} J.,  2019a, \mn@doi [\mnras] {10.1093/mnras/sty2913}, \href {https://ui.adsabs.harvard.edu/abs/2019MNRAS.483.1314B} {483, 1314}

\bibitem[\protect\citeauthoryear{{Buck}, {Dutton}  \& {Macci{\`o}}}{{Buck} et~al.}{2019b}]{Buck2019a}
{Buck} T.,  {Dutton} A.~A.,   {Macci{\`o}} A.~V.,  2019b, \mn@doi [\mnras] {10.1093/mnras/stz969}, \href {https://ui.adsabs.harvard.edu/abs/2019MNRAS.486.1481B} {486, 1481}

\bibitem[\protect\citeauthoryear{{Buck}, {Ness}, {Obreja}, {Macci{\`o}}  \& {Dutton}}{{Buck} et~al.}{2019c}]{Buck2019b}
{Buck} T.,  {Ness} M.,  {Obreja} A.,  {Macci{\`o}} A.~V.,   {Dutton} A.~A.,  2019c, \mn@doi [\apj] {10.3847/1538-4357/aaffd0}, \href {https://ui.adsabs.harvard.edu/abs/2019ApJ...874...67B} {874, 67}

\bibitem[\protect\citeauthoryear{{Buck}, {Obreja}, {Macci{\`o}}, {Minchev}, {Dutton}  \& {Ostriker}}{{Buck} et~al.}{2020}]{Buck2020}
{Buck} T.,  {Obreja} A.,  {Macci{\`o}} A.~V.,  {Minchev} I.,  {Dutton} A.~A.,   {Ostriker} J.~P.,  2020, \mn@doi [\mnras] {10.1093/mnras/stz3241}, \href {https://ui.adsabs.harvard.edu/abs/2020MNRAS.491.3461B} {491, 3461}

\bibitem[\protect\citeauthoryear{{Buck}, {Rybizki}, {Buder}, {Obreja}, {Macci{\`o}}, {Pfrommer}, {Steinmetz}  \& {Ness}}{{Buck} et~al.}{2021}]{Buck2021}
{Buck} T.,  {Rybizki} J.,  {Buder} S.,  {Obreja} A.,  {Macci{\`o}} A.~V.,  {Pfrommer} C.,  {Steinmetz} M.,   {Ness} M.,  2021, \mn@doi [\mnras] {10.1093/mnras/stab2736}, \href {https://ui.adsabs.harvard.edu/abs/2021MNRAS.508.3365B} {508, 3365}

\bibitem[\protect\citeauthoryear{{Buck}, {Obreja}, {Ratcliffe}, {Lu}, {Minchev}  \& {Macci{\`o}}}{{Buck} et~al.}{2023}]{Buck2023}
{Buck} T.,  {Obreja} A.,  {Ratcliffe} B.,  {Lu} Y.,  {Minchev} I.,   {Macci{\`o}} A.~V.,  2023, \mn@doi [\mnras] {10.1093/mnras/stad1503}, \href {https://ui.adsabs.harvard.edu/abs/2023MNRAS.523.1565B} {523, 1565}

\bibitem[\protect\citeauthoryear{{Bundy} et~al.,}{{Bundy} et~al.}{2015}]{MaNGA}
{Bundy} K.,  et~al., 2015, \mn@doi [\apj] {10.1088/0004-637X/798/1/7}, \href {https://ui.adsabs.harvard.edu/abs/2015ApJ...798....7B} {798, 7}

\bibitem[\protect\citeauthoryear{{Calzetti} et~al.,}{{Calzetti} et~al.}{2015}]{Calzetti2015}
{Calzetti} D.,  et~al., 2015, \mn@doi [\aj] {10.1088/0004-6256/149/2/51}, \href {http://adsabs.harvard.edu/abs/2015AJ....149...51C} {149, 51}

\bibitem[\protect\citeauthoryear{{Conroy} et~al.,}{{Conroy} et~al.}{2022}]{Conroy2022}
{Conroy} C.,  et~al., 2022, \mn@doi [arXiv e-prints] {10.48550/arXiv.2204.02989}, \href {https://ui.adsabs.harvard.edu/abs/2022arXiv220402989C} {p. arXiv:2204.02989}

\bibitem[\protect\citeauthoryear{{Cui} et~al.,}{{Cui} et~al.}{2012}]{LAMOST}
{Cui} X.-Q.,  et~al., 2012, \mn@doi [Research in Astronomy and Astrophysics] {10.1088/1674-4527/12/9/003}, \href {https://ui.adsabs.harvard.edu/abs/2012RAA....12.1197C} {12, 1197}

\bibitem[\protect\citeauthoryear{{Dutton}, {Macci{\`o}}, {Buck}, {Dixon}, {Blank}  \& {Obreja}}{{Dutton} et~al.}{2019}]{Dutton2019}
{Dutton} A.~A.,  {Macci{\`o}} A.~V.,  {Buck} T.,  {Dixon} K.~L.,  {Blank} M.,   {Obreja} A.,  2019, \mn@doi [\mnras] {10.1093/mnras/stz889}, \href {http://adsabs.harvard.edu/abs/2019MNRAS.tmp..863D} {}

\bibitem[\protect\citeauthoryear{{Dutton}, {Buck}, {Macci{\`o}}, {Dixon}, {Blank}  \& {Obreja}}{{Dutton} et~al.}{2020}]{Dutton2020}
{Dutton} A.~A.,  {Buck} T.,  {Macci{\`o}} A.~V.,  {Dixon} K.~L.,  {Blank} M.,   {Obreja} A.,  2020, \mn@doi [\mnras] {10.1093/mnras/staa3028}, \href {https://ui.adsabs.harvard.edu/abs/2020MNRAS.499.2648D} {499, 2648}

\bibitem[\protect\citeauthoryear{{El-Badry}, {Wetzel}, {Geha}, {Hopkins}, {Kere{\v{s}}}, {Chan}  \& {Faucher-Gigu{\`e}re}}{{El-Badry} et~al.}{2016}]{Elbadry2016}
{El-Badry} K.,  {Wetzel} A.,  {Geha} M.,  {Hopkins} P.~F.,  {Kere{\v{s}}} D.,  {Chan} T.~K.,   {Faucher-Gigu{\`e}re} C.-A.,  2016, \mn@doi [\apj] {10.3847/0004-637X/820/2/131}, \href {https://ui.adsabs.harvard.edu/abs/2016ApJ...820..131E} {820, 131}

\bibitem[\protect\citeauthoryear{{Frankel}, {Rix}, {Ting}, {Ness}  \& {Hogg}}{{Frankel} et~al.}{2018}]{Frankel2018}
{Frankel} N.,  {Rix} H.-W.,  {Ting} Y.-S.,  {Ness} M.,   {Hogg} D.~W.,  2018, \mn@doi [\apj] {10.3847/1538-4357/aadba5}, \href {https://ui.adsabs.harvard.edu/abs/2018ApJ...865...96F} {865, 96}

\bibitem[\protect\citeauthoryear{{Frankel}, {Sanders}, {Rix}, {Ting}  \& {Ness}}{{Frankel} et~al.}{2019}]{Frankel2019}
{Frankel} N.,  {Sanders} J.,  {Rix} H.-W.,  {Ting} Y.-S.,   {Ness} M.,  2019, \mn@doi [\apj] {10.3847/1538-4357/ab4254}, \href {https://ui.adsabs.harvard.edu/abs/2019ApJ...884...99F} {884, 99}

\bibitem[\protect\citeauthoryear{{Frankel}, {Sanders}, {Ting}  \& {Rix}}{{Frankel} et~al.}{2020}]{Frankel2020}
{Frankel} N.,  {Sanders} J.,  {Ting} Y.-S.,   {Rix} H.-W.,  2020, \mn@doi [\apj] {10.3847/1538-4357/ab910c}, \href {https://ui.adsabs.harvard.edu/abs/2020ApJ...896...15F} {896, 15}

\bibitem[\protect\citeauthoryear{{Gaia Collaboration} et~al.,}{{Gaia Collaboration} et~al.}{2016}]{gaia2016}
{Gaia Collaboration} et~al., 2016, \mn@doi [\aap] {10.1051/0004-6361/201629272}, \href {https://ui.adsabs.harvard.edu/abs/2016A&A...595A...1G} {595, A1}

\bibitem[\protect\citeauthoryear{{Gaia Collaboration} et~al.,}{{Gaia Collaboration} et~al.}{2021}]{gaiaLMC2021}
{Gaia Collaboration} et~al., 2021, \mn@doi [\aap] {10.1051/0004-6361/202039588}, \href {https://ui.adsabs.harvard.edu/abs/2021A&A...649A...7G} {649, A7}

\bibitem[\protect\citeauthoryear{{Gardner} et~al.,}{{Gardner} et~al.}{2006}]{JWST}
{Gardner} J.~P.,  et~al., 2006, \mn@doi [\ssr] {10.1007/s11214-006-8315-7}, \href {https://ui.adsabs.harvard.edu/abs/2006SSRv..123..485G} {123, 485}

\bibitem[\protect\citeauthoryear{{Goddard} et~al.,}{{Goddard} et~al.}{2017}]{Goddard2017}
{Goddard} D.,  et~al., 2017, \mn@doi [\mnras] {10.1093/mnras/stw3371}, \href {https://ui.adsabs.harvard.edu/abs/2017MNRAS.466.4731G} {466, 4731}

\bibitem[\protect\citeauthoryear{{Grasha} et~al.,}{{Grasha} et~al.}{2017}]{Grasha2017}
{Grasha} K.,  et~al., 2017, \mn@doi [\apj] {10.3847/1538-4357/aa6f15}, \href {http://adsabs.harvard.edu/abs/2017ApJ...840..113G} {840, 113}

\bibitem[\protect\citeauthoryear{{Grenon}}{{Grenon}}{1972}]{Grenon1972}
{Grenon} M.,  1972, in {Cayrel de Strobel} G.,  {Delplace} A.~M.,  eds, IAU Colloq. 17: Age des Etoiles. p.~55

\bibitem[\protect\citeauthoryear{{Grenon}}{{Grenon}}{1989}]{Grenon1989}
{Grenon} M.,  1989, \mn@doi [\apss] {10.1007/BF00646341}, \href {https://ui.adsabs.harvard.edu/abs/1989Ap&SS.156...29G} {156, 29}

\bibitem[\protect\citeauthoryear{{Haardt} \& {Madau}}{{Haardt} \& {Madau}}{2005}]{Haardt2005}
{Haardt} F.,  {Madau} P.,  2005, \mn@doi [unpublished] {10.1086/177035}, \href {http://adsabs.harvard.edu/cgi-bin/nph-bib_query?bibcode=1996ApJ...461...20H&db_key=AST} {}

\bibitem[\protect\citeauthoryear{{Hilmi} et~al.,}{{Hilmi} et~al.}{2020}]{Hilmi2020}
{Hilmi} T.,  et~al., 2020, \mn@doi [\mnras] {10.1093/mnras/staa1934}, \href {https://ui.adsabs.harvard.edu/abs/2020MNRAS.497..933H} {497, 933}

\bibitem[\protect\citeauthoryear{{Ho}, {Kudritzki}, {Kewley}, {Zahid}, {Dopita}, {Bresolin}  \& {Rupke}}{{Ho} et~al.}{2015}]{Ho2015}
{Ho} I.~T.,  {Kudritzki} R.-P.,  {Kewley} L.~J.,  {Zahid} H.~J.,  {Dopita} M.~A.,  {Bresolin} F.,   {Rupke} D. S.~N.,  2015, \mn@doi [\mnras] {10.1093/mnras/stv067}, \href {https://ui.adsabs.harvard.edu/abs/2015MNRAS.448.2030H} {448, 2030}

\bibitem[\protect\citeauthoryear{Hunter}{Hunter}{2007}]{matplotlib}
Hunter J.~D.,  2007, \mn@doi [Computing In Science \& Engineering] {10.1109/MCSE.2007.55}, 9, 90

\bibitem[\protect\citeauthoryear{Jones, Oliphant, Peterson  et~al.}{Jones et~al.}{01  }]{scipy}
Jones E.,  Oliphant T.,  Peterson P.,   et~al., 2001--, {SciPy}: Open source scientific tools for {Python}, \url {http://www.scipy.org/}

\bibitem[\protect\citeauthoryear{{Kaufmann}, {Wheeler}  \& {Bullock}}{{Kaufmann} et~al.}{2007}]{Kaufmann2007}
{Kaufmann} T.,  {Wheeler} C.,   {Bullock} J.~S.,  2007, \mn@doi [\mnras] {10.1111/j.1365-2966.2007.12436.x}, \href {https://ui.adsabs.harvard.edu/abs/2007MNRAS.382.1187K} {382, 1187}

\bibitem[\protect\citeauthoryear{Kluyver et~al.,}{Kluyver et~al.}{2016}]{jupyter}
Kluyver T.,  et~al., 2016, in Loizides F.,  Schmidt B.,  eds, Positioning and Power in Academic Publishing: Players, Agents and Agendas. pp 87 -- 90

\bibitem[\protect\citeauthoryear{{Knollmann} \& {Knebe}}{{Knollmann} \& {Knebe}}{2009}]{Knollmann2009}
{Knollmann} S.~R.,  {Knebe} A.,  2009, \mn@doi [\apjs] {10.1088/0067-0049/182/2/608}, \href {http://adsabs.harvard.edu/abs/2009ApJS..182..608K} {182, 608}

\bibitem[\protect\citeauthoryear{{Kobayashi}}{{Kobayashi}}{2004}]{Kobayashi2004}
{Kobayashi} C.,  2004, \mn@doi [\mnras] {10.1111/j.1365-2966.2004.07258.x}, \href {https://ui.adsabs.harvard.edu/abs/2004MNRAS.347..740K} {347, 740}

\bibitem[\protect\citeauthoryear{{Koleva}, {Prugniel}, {De Rijcke}  \& {Zeilinger}}{{Koleva} et~al.}{2011}]{Koleva2011}
{Koleva} M.,  {Prugniel} P.,  {De Rijcke} S.,   {Zeilinger} W.~W.,  2011, \mn@doi [\mnras] {10.1111/j.1365-2966.2011.19057.x}, \href {https://ui.adsabs.harvard.edu/abs/2011MNRAS.417.1643K} {417, 1643}

\bibitem[\protect\citeauthoryear{{Kollmeier} et~al.,}{{Kollmeier} et~al.}{2017}]{Kollmeier2017}
{Kollmeier} J.~A.,  et~al., 2017, \mn@doi [arXiv e-prints] {10.48550/arXiv.1711.03234}, \href {https://ui.adsabs.harvard.edu/abs/2017arXiv171103234K} {p. arXiv:1711.03234}

\bibitem[\protect\citeauthoryear{{Larson}}{{Larson}}{1974}]{Larson1974}
{Larson} R.~B.,  1974, \mn@doi [\mnras] {10.1093/mnras/166.3.585}, \href {https://ui.adsabs.harvard.edu/abs/1974MNRAS.166..585L} {166, 585}

\bibitem[\protect\citeauthoryear{{Lelli}, {McGaugh}  \& {Schombert}}{{Lelli} et~al.}{2016}]{Lelli2016}
{Lelli} F.,  {McGaugh} S.~S.,   {Schombert} J.~M.,  2016, \mn@doi [\aj] {10.3847/0004-6256/152/6/157}, \href {https://ui.adsabs.harvard.edu/abs/2016AJ....152..157L} {152, 157}

\bibitem[\protect\citeauthoryear{{Lian} et~al.,}{{Lian} et~al.}{2022}]{Lian2022}
{Lian} J.,  et~al., 2022, \mn@doi [\mnras] {10.1093/mnras/stac479}, \href {https://ui.adsabs.harvard.edu/abs/2022MNRAS.511.5639L} {511, 5639}

\bibitem[\protect\citeauthoryear{{Lu}, {Minchev}, {Buck}, {Khoperskov}, {Steinmetz}, {Libeskind}, {Cescutti}  \& {Freeman}}{{Lu} et~al.}{2022a}]{Lu2022_rb}
{Lu} Y.,  {Minchev} I.,  {Buck} T.,  {Khoperskov} S.,  {Steinmetz} M.,  {Libeskind} N.,  {Cescutti} G.,   {Freeman} K.~C.,  2022a, \mn@doi [arXiv e-prints] {10.48550/arXiv.2212.04515}, \href {https://ui.adsabs.harvard.edu/abs/2022arXiv221204515L} {p. arXiv:2212.04515}

\bibitem[\protect\citeauthoryear{{Lu}, {Ness}, {Buck}  \& {Carr}}{{Lu} et~al.}{2022b}]{Lu2022_turning}
{Lu} Y.~L.,  {Ness} M.~K.,  {Buck} T.,   {Carr} C.,  2022b, \mn@doi [\mnras] {10.1093/mnras/stac780}, \href {https://ui.adsabs.harvard.edu/abs/2022MNRAS.512.4697L} {512, 4697}

\bibitem[\protect\citeauthoryear{{Lu}, {Buck}, {Minchev}  \& {Ness}}{{Lu} et~al.}{2022c}]{Lu2022_rblim}
{Lu} Y.,  {Buck} T.,  {Minchev} I.,   {Ness} M.~K.,  2022c, \mn@doi [\mnras] {10.1093/mnrasl/slac065}, \href {https://ui.adsabs.harvard.edu/abs/2022MNRAS.515L..34L} {515, L34}

\bibitem[\protect\citeauthoryear{{Ma}, {Hopkins}, {Wetzel}, {Kirby}, {Angl{\'e}s-Alc{\'a}zar}, {Faucher-Gigu{\`e}re}, {Kere{\v{s}}}  \& {Quataert}}{{Ma} et~al.}{2017}]{Ma2017}
{Ma} X.,  {Hopkins} P.~F.,  {Wetzel} A.~R.,  {Kirby} E.~N.,  {Angl{\'e}s-Alc{\'a}zar} D.,  {Faucher-Gigu{\`e}re} C.-A.,  {Kere{\v{s}}} D.,   {Quataert} E.,  2017, \mn@doi [\mnras] {10.1093/mnras/stx273}, \href {https://ui.adsabs.harvard.edu/abs/2017MNRAS.467.2430M} {467, 2430}

\bibitem[\protect\citeauthoryear{{Macci{\`o}}, {Ali-Dib}, {Vulanovic}, {Al Noori}, {Walter}, {Krieger}  \& {Buck}}{{Macci{\`o}} et~al.}{2022}]{Maccio2022}
{Macci{\`o}} A.~V.,  {Ali-Dib} M.,  {Vulanovic} P.,  {Al Noori} H.,  {Walter} F.,  {Krieger} N.,   {Buck} T.,  2022, \mn@doi [\mnras] {10.1093/mnras/stac482}, \href {https://ui.adsabs.harvard.edu/abs/2022MNRAS.512.2135M} {512, 2135}

\bibitem[\protect\citeauthoryear{{Majewski} et~al.,}{{Majewski} et~al.}{2017}]{APOGEE2017}
{Majewski} S.~R.,  et~al., 2017, \mn@doi [\aj] {10.3847/1538-3881/aa784d}, \href {https://ui.adsabs.harvard.edu/abs/2017AJ....154...94M} {154, 94}

\bibitem[\protect\citeauthoryear{{Marinacci} et~al.,}{{Marinacci} et~al.}{2018}]{Marinacci2018_TNG}
{Marinacci} F.,  et~al., 2018, \mn@doi [\mnras] {10.1093/mnras/sty2206}, \href {https://ui.adsabs.harvard.edu/abs/2018MNRAS.480.5113M} {480, 5113}

\bibitem[\protect\citeauthoryear{{McKee} \& {Ostriker}}{{McKee} \& {Ostriker}}{1977}]{McKee1977}
{McKee} C.~F.,  {Ostriker} J.~P.,  1977, \mn@doi [\apj] {10.1086/155667}, \href {https://ui.adsabs.harvard.edu/abs/1977ApJ...218..148M} {218, 148}

\bibitem[\protect\citeauthoryear{{Mercado} et~al.,}{{Mercado} et~al.}{2021}]{Mercado2021}
{Mercado} F.~J.,  et~al., 2021, \mn@doi [\mnras] {10.1093/mnras/staa3958}, \href {https://ui.adsabs.harvard.edu/abs/2021MNRAS.501.5121M} {501, 5121}

\bibitem[\protect\citeauthoryear{{Minchev} \& {Famaey}}{{Minchev} \& {Famaey}}{2010}]{Minchev2010}
{Minchev} I.,  {Famaey} B.,  2010, \mn@doi [\apj] {10.1088/0004-637X/722/1/112}, \href {https://ui.adsabs.harvard.edu/abs/2010ApJ...722..112M} {722, 112}

\bibitem[\protect\citeauthoryear{{Minchev} \& {Quillen}}{{Minchev} \& {Quillen}}{2006}]{Minchev2006}
{Minchev} I.,  {Quillen} A.~C.,  2006, \mn@doi [\mnras] {10.1111/j.1365-2966.2006.10129.x}, \href {https://ui.adsabs.harvard.edu/abs/2006MNRAS.368..623M} {368, 623}

\bibitem[\protect\citeauthoryear{{Minchev}, {Famaey}, {Quillen}, {Dehnen}, {Martig}  \& {Siebert}}{{Minchev} et~al.}{2012}]{Minchev2012}
{Minchev} I.,  {Famaey} B.,  {Quillen} A.~C.,  {Dehnen} W.,  {Martig} M.,   {Siebert} A.,  2012, \mn@doi [\aap] {10.1051/0004-6361/201219714}, \href {https://ui.adsabs.harvard.edu/abs/2012A&A...548A.127M} {548, A127}

\bibitem[\protect\citeauthoryear{{Minchev}, {Chiappini}  \& {Martig}}{{Minchev} et~al.}{2013}]{Minchev2013}
{Minchev} I.,  {Chiappini} C.,   {Martig} M.,  2013, \mn@doi [\aap] {10.1051/0004-6361/201220189}, \href {https://ui.adsabs.harvard.edu/abs/2013A&A...558A...9M} {558, A9}

\bibitem[\protect\citeauthoryear{{Minchev} et~al.,}{{Minchev} et~al.}{2018}]{Minchev2018}
{Minchev} I.,  et~al., 2018, \mn@doi [\mnras] {10.1093/mnras/sty2033}, \href {https://ui.adsabs.harvard.edu/abs/2018MNRAS.481.1645M} {481, 1645}

\bibitem[\protect\citeauthoryear{{Naiman} et~al.,}{{Naiman} et~al.}{2018}]{Naiman2018_TNG}
{Naiman} J.~P.,  et~al., 2018, \mn@doi [\mnras] {10.1093/mnras/sty618}, \href {https://ui.adsabs.harvard.edu/abs/2018MNRAS.477.1206N} {477, 1206}

\bibitem[\protect\citeauthoryear{{Nanni} et~al.,}{{Nanni} et~al.}{2023}]{Nanni2023}
{Nanni} L.,  et~al., 2023, \mn@doi [arXiv e-prints] {10.48550/arXiv.2309.14257}, \href {https://ui.adsabs.harvard.edu/abs/2023arXiv230914257N} {p. arXiv:2309.14257}

\bibitem[\protect\citeauthoryear{{Nelson} et~al.,}{{Nelson} et~al.}{2018}]{Nelson2018_TNG}
{Nelson} D.,  et~al., 2018, \mn@doi [\mnras] {10.1093/mnras/stx3040}, \href {https://ui.adsabs.harvard.edu/abs/2018MNRAS.475..624N} {475, 624}

\bibitem[\protect\citeauthoryear{{Nelson} et~al.,}{{Nelson} et~al.}{2019a}]{Nelson2019a_TNG}
{Nelson} D.,  et~al., 2019a, \mn@doi [Computational Astrophysics and Cosmology] {10.1186/s40668-019-0028-x}, \href {https://ui.adsabs.harvard.edu/abs/2019ComAC...6....2N} {6, 2}

\bibitem[\protect\citeauthoryear{{Nelson} et~al.,}{{Nelson} et~al.}{2019b}]{Nelson2019}
{Nelson} D.,  et~al., 2019b, \mn@doi [\mnras] {10.1093/mnras/stz2306}, \href {https://ui.adsabs.harvard.edu/abs/2019MNRAS.490.3234N} {490, 3234}

\bibitem[\protect\citeauthoryear{{Nidever} et~al.,}{{Nidever} et~al.}{2020}]{Nidever2020}
{Nidever} D.~L.,  et~al., 2020, \mn@doi [\apj] {10.3847/1538-4357/ab7305}, \href {https://ui.adsabs.harvard.edu/abs/2020ApJ...895...88N} {895, 88}

\bibitem[\protect\citeauthoryear{{Nidever} et~al.,}{{Nidever} et~al.}{2023}]{Nidever2023}
{Nidever} D.~L.,  et~al., 2023, \mn@doi [arXiv e-prints] {10.48550/arXiv.2306.04688}, \href {https://ui.adsabs.harvard.edu/abs/2023arXiv230604688N} {p. arXiv:2306.04688}

\bibitem[\protect\citeauthoryear{{Obreja} et~al.,}{{Obreja} et~al.}{2019}]{Obreja2019}
{Obreja} A.,  et~al., 2019, \mn@doi [\mnras] {10.1093/mnras/stz1563}, \href {https://ui.adsabs.harvard.edu/abs/2019MNRAS.487.4424O} {487, 4424}

\bibitem[\protect\citeauthoryear{{Obreja}, {Buck}  \& {Macci{\`o}}}{{Obreja} et~al.}{2022}]{Obreja2022}
{Obreja} A.,  {Buck} T.,   {Macci{\`o}} A.~V.,  2022, \mn@doi [\aap] {10.1051/0004-6361/202140983}, \href {https://ui.adsabs.harvard.edu/abs/2022A&A...657A..15O} {657, A15}

\bibitem[\protect\citeauthoryear{{Parikh}, {Thomas}, {Maraston}, {Westfall}, {Andrews}, {Boardman}, {Drory}  \& {Oyarzun}}{{Parikh} et~al.}{2021}]{Parikh2021}
{Parikh} T.,  {Thomas} D.,  {Maraston} C.,  {Westfall} K.~B.,  {Andrews} B.~H.,  {Boardman} N.~F.,  {Drory} N.,   {Oyarzun} G.,  2021, \mn@doi [\mnras] {10.1093/mnras/stab449}, \href {https://ui.adsabs.harvard.edu/abs/2021MNRAS.502.5508P} {502, 5508}

\bibitem[\protect\citeauthoryear{{Penzo}, {Macci{\`o}}, {Casarini}, {Stinson}  \& {Wadsley}}{{Penzo} et~al.}{2014}]{Penzo2014}
{Penzo} C.,  {Macci{\`o}} A.~V.,  {Casarini} L.,  {Stinson} G.~S.,   {Wadsley} J.,  2014, \mn@doi [\mnras] {10.1093/mnras/stu857}, \href {http://adsabs.harvard.edu/abs/2014MNRAS.442..176P} {442, 176}

\bibitem[\protect\citeauthoryear{P\'erez \& Granger}{P\'erez \& Granger}{2007}]{ipython}
P\'erez F.,  Granger B.~E.,  2007, \mn@doi [Computing in Science and Engineering] {10.1109/MCSE.2007.53}, 9, 21

\bibitem[\protect\citeauthoryear{{Pillepich} et~al.,}{{Pillepich} et~al.}{2018}]{Pillepich2018_TNG}
{Pillepich} A.,  et~al., 2018, \mn@doi [\mnras] {10.1093/mnras/stx3112}, \href {https://ui.adsabs.harvard.edu/abs/2018MNRAS.475..648P} {475, 648}

\bibitem[\protect\citeauthoryear{{Pillepich} et~al.,}{{Pillepich} et~al.}{2019}]{Pillepich2019}
{Pillepich} A.,  et~al., 2019, \mn@doi [\mnras] {10.1093/mnras/stz2338}, \href {https://ui.adsabs.harvard.edu/abs/2019MNRAS.490.3196P} {490, 3196}

\bibitem[\protect\citeauthoryear{{Pilyugin}, {Tautvai{\v{s}}ien{\.{e}}}  \& {Lara-L{\'o}pez}}{{Pilyugin} et~al.}{2023}]{Pilyugin2023}
{Pilyugin} L.~S.,  {Tautvai{\v{s}}ien{\.{e}}} G.,   {Lara-L{\'o}pez} M.~A.,  2023, \mn@doi [\aap] {10.1051/0004-6361/202346503}, \href {https://ui.adsabs.harvard.edu/abs/2023A&A...676A..57P} {676, A57}

\bibitem[\protect\citeauthoryear{{Planck Collaboration} et~al.,}{{Planck Collaboration} et~al.}{2014}]{Planck}
{Planck Collaboration} et~al., 2014, \mn@doi [\aap] {10.1051/0004-6361/201321591}, \href {http://adsabs.harvard.edu/abs/2014A%26A...571A..16P} {571, A16}

\bibitem[\protect\citeauthoryear{{Pontzen}, {Ro{\v s}kar}, {Stinson}, {Woods}, {Reed}, {Coles}  \& {Quinn}}{{Pontzen} et~al.}{2013}]{pynbody}
{Pontzen} A.,  {Ro{\v s}kar} R.,  {Stinson} G.~S.,  {Woods} R.,  {Reed} D.~M.,  {Coles} J.,   {Quinn} T.~R.,  2013, {pynbody: Astrophysics Simulation Analysis for Python}

\bibitem[\protect\citeauthoryear{{Povick} et~al.,}{{Povick} et~al.}{2023a}]{Povick2023}
{Povick} J.~T.,  et~al., 2023a, \mn@doi [arXiv e-prints] {10.48550/arXiv.2306.06348}, \href {https://ui.adsabs.harvard.edu/abs/2023arXiv230606348P} {p. arXiv:2306.06348}

\bibitem[\protect\citeauthoryear{{Povick} et~al.,}{{Povick} et~al.}{2023b}]{Povick2023b}
{Povick} J.~T.,  et~al., 2023b, \mn@doi [arXiv e-prints] {10.48550/arXiv.2310.14299}, \href {https://ui.adsabs.harvard.edu/abs/2023arXiv231014299P} {p. arXiv:2310.14299}

\bibitem[\protect\citeauthoryear{{Quillen}, {Minchev}, {Bland-Hawthorn}  \& {Haywood}}{{Quillen} et~al.}{2009}]{Quillen2009}
{Quillen} A.~C.,  {Minchev} I.,  {Bland-Hawthorn} J.,   {Haywood} M.,  2009, \mn@doi [\mnras] {10.1111/j.1365-2966.2009.15054.x}, \href {https://ui.adsabs.harvard.edu/abs/2009MNRAS.397.1599Q} {397, 1599}

\bibitem[\protect\citeauthoryear{{Ratcliffe}, {Minchev}, {Cescutti}, {Spitoni}, {J{\"o}nsson}, {Anders}, {Queiroz}  \& {Steinmetz}}{{Ratcliffe} et~al.}{2023a}]{Ratcliffe2023_2}
{Ratcliffe} B.,  {Minchev} I.,  {Cescutti} G.,  {Spitoni} E.,  {J{\"o}nsson} H.,  {Anders} F.,  {Queiroz} A.,   {Steinmetz} M.,  2023a, \mn@doi [arXiv e-prints] {10.48550/arXiv.2307.11159}, \href {https://ui.adsabs.harvard.edu/abs/2023arXiv230711159R} {p. arXiv:2307.11159}

\bibitem[\protect\citeauthoryear{{Ratcliffe} et~al.,}{{Ratcliffe} et~al.}{2023b}]{Ratcliffe2023}
{Ratcliffe} B.,  et~al., 2023b, \mn@doi [\mnras] {10.1093/mnras/stad1573}, \href {https://ui.adsabs.harvard.edu/abs/2023MNRAS.525.2208R} {525, 2208}

\bibitem[\protect\citeauthoryear{{Roig}, {Blanton}  \& {Yan}}{{Roig} et~al.}{2015}]{Roig2015}
{Roig} B.,  {Blanton} M.~R.,   {Yan} R.,  2015, \mn@doi [\apj] {10.1088/0004-637X/808/1/26}, \href {https://ui.adsabs.harvard.edu/abs/2015ApJ...808...26R} {808, 26}

\bibitem[\protect\citeauthoryear{{Ro{\v{s}}kar}, {Debattista}, {Stinson}, {Quinn}, {Kaufmann}  \& {Wadsley}}{{Ro{\v{s}}kar} et~al.}{2008}]{Roskar2008}
{Ro{\v{s}}kar} R.,  {Debattista} V.~P.,  {Stinson} G.~S.,  {Quinn} T.~R.,  {Kaufmann} T.,   {Wadsley} J.,  2008, \mn@doi [\apjl] {10.1086/586734}, \href {https://ui.adsabs.harvard.edu/abs/2008ApJ...675L..65R} {675, L65}

\bibitem[\protect\citeauthoryear{{S{\'a}nchez-Bl{\'a}zquez} et~al.,}{{S{\'a}nchez-Bl{\'a}zquez} et~al.}{2014}]{sanchezblazquez2014}
{S{\'a}nchez-Bl{\'a}zquez} P.,  et~al., 2014, \mn@doi [\aap] {10.1051/0004-6361/201423635}, \href {https://ui.adsabs.harvard.edu/abs/2014A&A...570A...6S} {570, A6}

\bibitem[\protect\citeauthoryear{{Schroyen}, {De Rijcke}, {Koleva}, {Cloet-Osselaer}  \& {Vandenbroucke}}{{Schroyen} et~al.}{2013}]{Schroyen2013}
{Schroyen} J.,  {De Rijcke} S.,  {Koleva} M.,  {Cloet-Osselaer} A.,   {Vandenbroucke} B.,  2013, \mn@doi [\mnras] {10.1093/mnras/stt1084}, \href {https://ui.adsabs.harvard.edu/abs/2013MNRAS.434..888S} {434, 888}

\bibitem[\protect\citeauthoryear{{Sellwood} \& {Binney}}{{Sellwood} \& {Binney}}{2002}]{Sellwood2002}
{Sellwood} J.~A.,  {Binney} J.~J.,  2002, \mn@doi [\mnras] {10.1046/j.1365-8711.2002.05806.x}, \href {https://ui.adsabs.harvard.edu/abs/2002MNRAS.336..785S} {336, 785}

\bibitem[\protect\citeauthoryear{{S{\'e}rsic}}{{S{\'e}rsic}}{1963}]{sersic1963}
{S{\'e}rsic} J.~L.,  1963, Boletin de la Asociacion Argentina de Astronomia La Plata Argentina, \href {https://ui.adsabs.harvard.edu/abs/1963BAAA....6...41S} {6, 41}

\bibitem[\protect\citeauthoryear{{Sestito} et~al.,}{{Sestito} et~al.}{2021}]{Sestito2021}
{Sestito} F.,  et~al., 2021, \mn@doi [\mnras] {10.1093/mnras/staa3479}, \href {https://ui.adsabs.harvard.edu/abs/2021MNRAS.500.3750S} {500, 3750}

\bibitem[\protect\citeauthoryear{{Shen}, {Wadsley}  \& {Stinson}}{{Shen} et~al.}{2010}]{Shen2010}
{Shen} S.,  {Wadsley} J.,   {Stinson} G.,  2010, \mn@doi [\mnras] {10.1111/j.1365-2966.2010.17047.x}, \href {http://adsabs.harvard.edu/abs/2010MNRAS.407.1581S} {407, 1581}

\bibitem[\protect\citeauthoryear{{Shipp} et~al.,}{{Shipp} et~al.}{2021}]{Shipp2021}
{Shipp} N.,  et~al., 2021, \mn@doi [\apj] {10.3847/1538-4357/ac2e93}, \href {https://ui.adsabs.harvard.edu/abs/2021ApJ...923..149S} {923, 149}

\bibitem[\protect\citeauthoryear{Smith \& Lang}{Smith \& Lang}{2019}]{ytree}
Smith B.~D.,  Lang M.,  2019, \mn@doi [Journal of Open Source Software] {10.21105/joss.01881}, 4, 1881

\bibitem[\protect\citeauthoryear{{Spergel} et~al.,}{{Spergel} et~al.}{2015}]{Roman}
{Spergel} D.,  et~al., 2015, \mn@doi [arXiv e-prints] {10.48550/arXiv.1503.03757}, \href {https://ui.adsabs.harvard.edu/abs/2015arXiv150303757S} {p. arXiv:1503.03757}

\bibitem[\protect\citeauthoryear{{Spolaor}, {Proctor}, {Forbes}  \& {Couch}}{{Spolaor} et~al.}{2009}]{Spolaor2009}
{Spolaor} M.,  {Proctor} R.~N.,  {Forbes} D.~A.,   {Couch} W.~J.,  2009, \mn@doi [\apjl] {10.1088/0004-637X/691/2/L138}, \href {https://ui.adsabs.harvard.edu/abs/2009ApJ...691L.138S} {691, L138}

\bibitem[\protect\citeauthoryear{{Springel} et~al.,}{{Springel} et~al.}{2018}]{Springel2018_TNG}
{Springel} V.,  et~al., 2018, \mn@doi [\mnras] {10.1093/mnras/stx3304}, \href {https://ui.adsabs.harvard.edu/abs/2018MNRAS.475..676S} {475, 676}

\bibitem[\protect\citeauthoryear{{Stinson}, {Seth}, {Katz}, {Wadsley}, {Governato}  \& {Quinn}}{{Stinson} et~al.}{2006}]{Stinson2006}
{Stinson} G.,  {Seth} A.,  {Katz} N.,  {Wadsley} J.,  {Governato} F.,   {Quinn} T.,  2006, \mn@doi [\mnras] {10.1111/j.1365-2966.2006.11097.x}, \href {http://esoads.eso.org/abs/2006MNRAS.373.1074S} {373, 1074}

\bibitem[\protect\citeauthoryear{{Stinson}, {Brook}, {Macci{\`o}}, {Wadsley}, {Quinn}  \& {Couchman}}{{Stinson} et~al.}{2013}]{Stinson2013}
{Stinson} G.~S.,  {Brook} C.,  {Macci{\`o}} A.~V.,  {Wadsley} J.,  {Quinn} T.~R.,   {Couchman} H.~M.~P.,  2013, \mn@doi [\mnras] {10.1093/mnras/sts028}, \href {http://adsabs.harvard.edu/abs/2013MNRAS.428..129S} {428, 129}

\bibitem[\protect\citeauthoryear{{Wadsley}, {Veeravalli}  \& {Couchman}}{{Wadsley} et~al.}{2008}]{Wadsley2008}
{Wadsley} J.~W.,  {Veeravalli} G.,   {Couchman} H.~M.~P.,  2008, \mn@doi [\mnras] {10.1111/j.1365-2966.2008.13260.x}, \href {http://adsabs.harvard.edu/abs/2008MNRAS.387..427W} {387, 427}

\bibitem[\protect\citeauthoryear{{Wadsley}, {Keller}  \& {Quinn}}{{Wadsley} et~al.}{2017}]{Wadsley2017}
{Wadsley} J.~W.,  {Keller} B.~W.,   {Quinn} T.~R.,  2017, \mn@doi [\mnras] {10.1093/mnras/stx1643}, \href {http://adsabs.harvard.edu/abs/2017MNRAS.471.2357W} {471, 2357}

\bibitem[\protect\citeauthoryear{Walt, Colbert  \& Varoquaux}{Walt et~al.}{2011}]{numpy}
Walt S. v.~d.,  Colbert S.~C.,   Varoquaux G.,  2011, \mn@doi [Computing in Science and Engg.] {10.1109/MCSE.2011.37}, 13, 22

\bibitem[\protect\citeauthoryear{{Walter}, {Brinks}, {de Blok}, {Bigiel}, {Kennicutt}, {Thornley}  \& {Leroy}}{{Walter} et~al.}{2008}]{Walter2008}
{Walter} F.,  {Brinks} E.,  {de Blok} W.~J.~G.,  {Bigiel} F.,  {Kennicutt} Robert~C. J.,  {Thornley} M.~D.,   {Leroy} A.,  2008, \mn@doi [\aj] {10.1088/0004-6256/136/6/2563}, \href {https://ui.adsabs.harvard.edu/abs/2008AJ....136.2563W} {136, 2563}

\bibitem[\protect\citeauthoryear{{Wang}, {Dutton}, {Stinson}, {Macci{\`o}}, {Penzo}, {Kang}, {Keller}  \& {Wadsley}}{{Wang} et~al.}{2015}]{Wang2015}
{Wang} L.,  {Dutton} A.~A.,  {Stinson} G.~S.,  {Macci{\`o}} A.~V.,  {Penzo} C.,  {Kang} X.,  {Keller} B.~W.,   {Wadsley} J.,  2015, \mn@doi [\mnras] {10.1093/mnras/stv1937}, \href {https://ui.adsabs.harvard.edu/abs/2015MNRAS.454...83W} {454, 83}

\bibitem[\protect\citeauthoryear{{Wang}, {Carrillo}, {Ness}  \& {Buck}}{{Wang} et~al.}{2023}]{Wang2023}
{Wang} K.,  {Carrillo} A.,  {Ness} M.~K.,   {Buck} T.,  2023, \mn@doi [arXiv e-prints] {10.48550/arXiv.2307.04724}, \href {https://ui.adsabs.harvard.edu/abs/2023arXiv230704724W} {p. arXiv:2307.04724}

\bibitem[\protect\citeauthoryear{{Xiang} \& {Rix}}{{Xiang} \& {Rix}}{2022}]{Xiang2022}
{Xiang} M.,  {Rix} H.-W.,  2022, \mn@doi [\nat] {10.1038/s41586-022-04496-5}, \href {https://ui.adsabs.harvard.edu/abs/2022Natur.603..599X} {603, 599}

\bibitem[\protect\citeauthoryear{{Xiang} et~al.,}{{Xiang} et~al.}{2019}]{Xiang2019}
{Xiang} M.,  et~al., 2019, \mn@doi [\apjs] {10.3847/1538-4365/ab5364}, \href {https://ui.adsabs.harvard.edu/abs/2019ApJS..245...34X} {245, 34}

\bibitem[\protect\citeauthoryear{{Zhang}, {Chen}, {Zhao}, {Bi}, {Zhang}  \& {Xue}}{{Zhang} et~al.}{2023}]{Zhang2023}
{Zhang} H.,  {Chen} Y.,  {Zhao} G.,  {Bi} S.,  {Zhang} X.,   {Xue} X.,  2023, \mn@doi [\mnras] {10.1093/mnras/stad348}, \href {https://ui.adsabs.harvard.edu/abs/2023MNRAS.520.4815Z} {520, 4815}

\bibitem[\protect\citeauthoryear{{Zheng} et~al.,}{{Zheng} et~al.}{2017}]{Zheng2017}
{Zheng} Z.,  et~al., 2017, \mn@doi [\mnras] {10.1093/mnras/stw3030}, \href {https://ui.adsabs.harvard.edu/abs/2017MNRAS.465.4572Z} {465, 4572}

\bibitem[\protect\citeauthoryear{{van der Marel} \& {Kallivayalil}}{{van der Marel} \& {Kallivayalil}}{2014}]{vanderMarel2014}
{van der Marel} R.~P.,  {Kallivayalil} N.,  2014, \mn@doi [\apj] {10.1088/0004-637X/781/2/121}, \href {https://ui.adsabs.harvard.edu/abs/2014ApJ...781..121V} {781, 121}

\makeatother
\end{thebibliography}




\appendix

\section{Azimuthally averaged stellar mass surface density profiles with best-fit models}
\begin{figure*}
	\includegraphics[width=\textwidth]{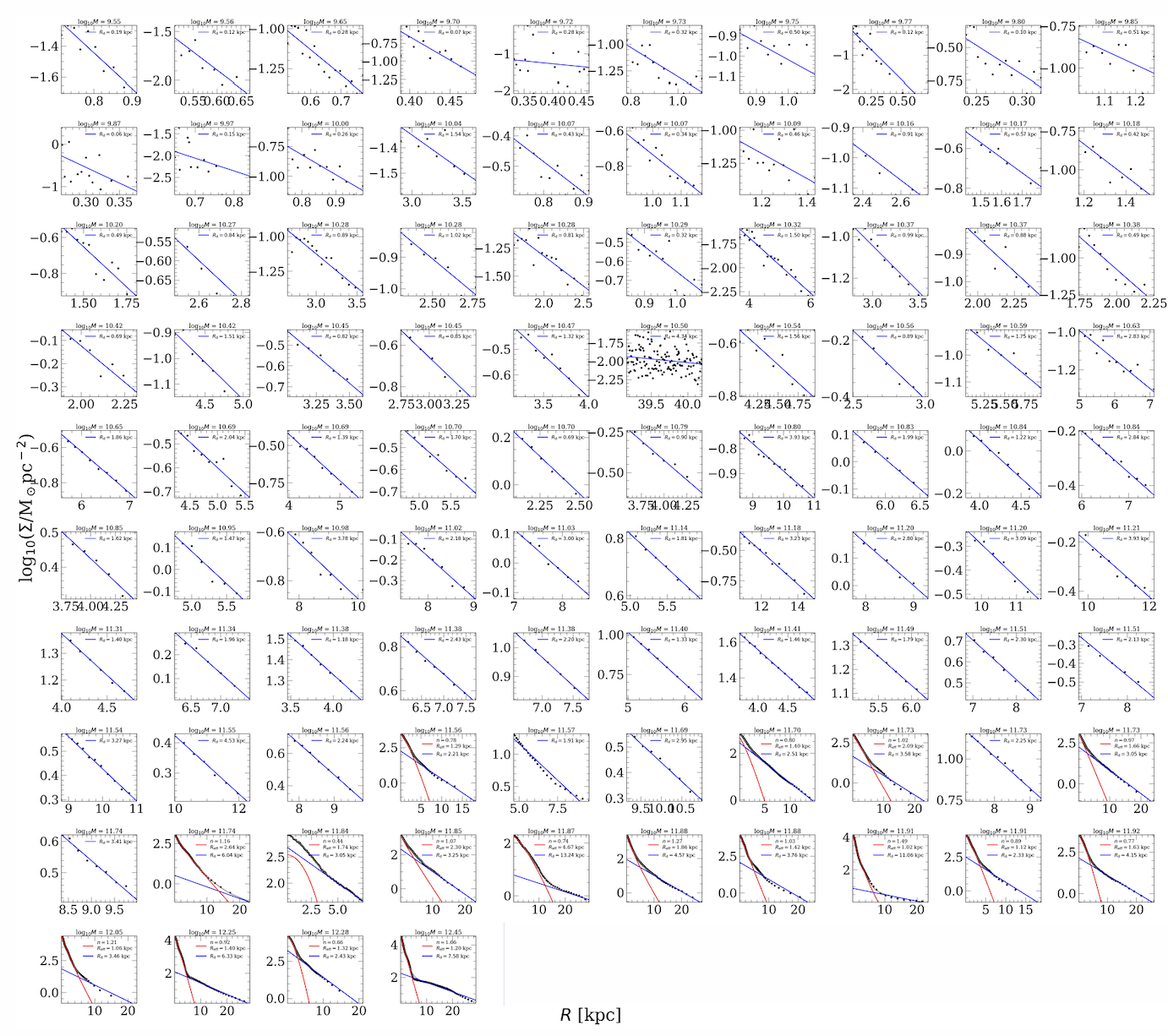}
    \caption{The best-fit model overlaying on the stellar mass density profile for the full radial range (stellar mass $> 1\times10^{10} M_\odot$) or the galaxy outskirt (stellar mass $< 1\times10^{10} M_\odot$) for each NIHAO galaxy.
    For galaxies with stellar mass $<$ 1$\times$10$^{10} M_\odot$, we fit only an expoential to the galaxy outskirts to match with what was done for SPARC \citep{Lelli2016}.
    For more massive galaxies, we simultaneously fitted a Sérsic \citep[red;][]{sersic1963} and an exponential profile (blue). 
    The Sérsic indix ($n$), effective radius ($R_{\rm eff}$), and galaxy scale length ($R_d$) from the best-fit models are shown in the legends.}
    \label{fig:A1}
\end{figure*}

Figure~\ref{fig:A1} shows the best-fit model overlaying on the stellar mass density profile for the full radial range (stellar mass $> 1\times10^{10} M_\odot$) or the galaxy outskirt (stellar mass $< 1\times10^{10} M_\odot$) for each NIHAO galaxy.

\section{Visualization of the face-on and edge-on view for the NIHAO galaxies}
\begin{figure*}
    \includegraphics[width=\textwidth]{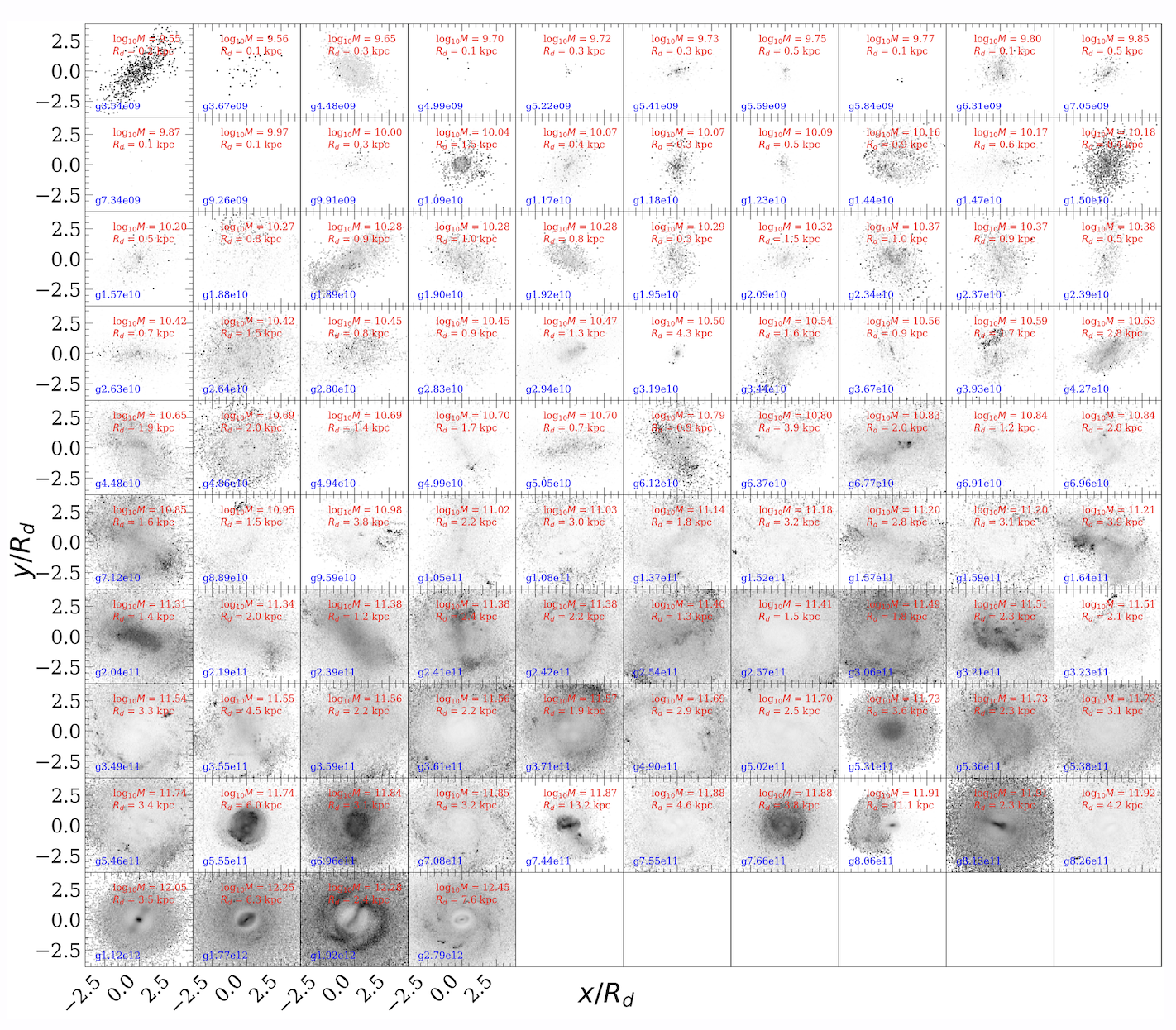}
    \caption{Face-on surface density maps for stars in the NIHAO galaxies ordered by mass (increasing from top left to bottom right).
    The positions are normalized by the scale length, $R_d$, of each galaxy to emphasize the morphology. 
    The galaxy name is shown in bottom left legends and the dark matter halo mass and $R_d$ of each galaxy are shown in the top right legends.}
    \label{fig:1}
\end{figure*}

\begin{figure*}
	\includegraphics[width=\textwidth]{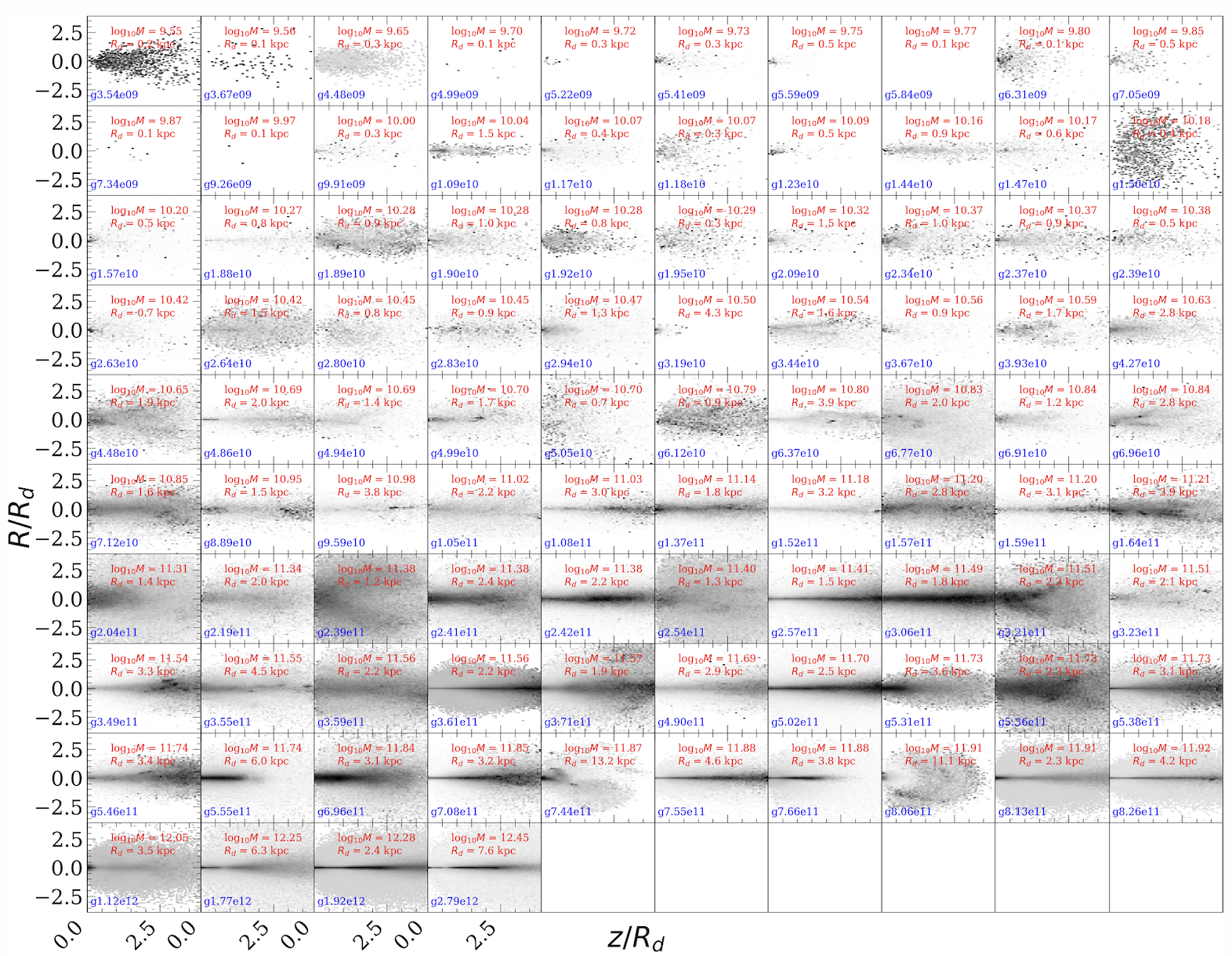}
    \caption{Same as figure~\ref{fig:1} but plotting radius, $R/R_d$, vs galactic height, $z/R_d$.}
    \label{fig:2}
\end{figure*}

Figure~\ref{fig:1} and Figure~\ref{fig:2} show the face-on ($x/R_d$ vs $y/R_d$) and edge-on ($R/R_d$ vs $z/R_d$) surface density profile for each galaxy, in which $R_d$ is the scale length of the galaxy today (see the previous section for details on how we measured $R_d$). 
The galaxy name is shown in the bottom left corner in blue. 
The mass of the dark matter halo increases from the top left corner to the bottom right. 

\bsp	
\label{lastpage}
\end{document}